\documentclass[aps,prb,twocolumn,nofootinbib,superscriptaddress]{revtex4-1}

\usepackage{graphicx}

\usepackage{color}
\usepackage{amsmath, amssymb,bm}

\usepackage{comment}
\newcommand{\1}{\mbox{1}\hspace{-0.25em}\mbox{l}}

\begin{document}


\title{Probing three-state Potts nematic fluctuations by ultrasound attenuation}


\author{Kazuhiro Kimura}
\email[]{E-mail address: kimura.kazuhiro.85n@st.kyoto-u.ac.jp}
\affiliation{Department of Physics, Kyoto University, Kyoto 606-8502, Japan}

\author{Manfred Sigrist} 
\affiliation{Institute for Theoretical Physics, ETH Zurich, 8093 Zurich, Switzerland}

\author{Norio Kawakami} 
\affiliation{Department of Physics, Kyoto University, Kyoto 606-8502, Japan}


\date{\today}

\begin{abstract}
Motivated by recent studies of three-state Potts nematic states in magic-angle twisted bilayer graphene and doped-Bi$_2$Se$_3$, we analyze the impact of critical nematic fluctuations on the low energy properties of phonons. 
In this study we propose how to identify the three-state Potts nematic fluctuations by ultrasound attenuation. 
The Gaussian fluctuation analysis shows that the Landau damping term becomes isotropic due to fluctuations of the $C_{3}$-breaking bond-order, and the nemato-elastic coupling is also shown to be isotropic. These two features lead to an isotropic divergence of the transverse sound attenuation coefficient and an isotropic lattice softening, in contrast to the case of the $C_4$-breaking bond-order, which shows strong anisotropy. Moreover, we use a mean-field approximation and discuss the impurity effects. The transition temperature takes its maximum near the filling of the van-Hove singularity, and the large density of states favors the nematic phase transition. It turns out that the phase transition is of weak first-order in the wide range of filling and, upon increasing the impurity scattering, the first-order transition line at low temperatures gradually shifts towards the second-order line, rendering the transition a weak first-order in a wider range of parameters. Furthermore, it is confirmed that the enhancement of the ultrasound attenuation coefficient will be clearly observed in experiments in the case of a weak first-order phase transition. 
\end{abstract}


\maketitle

\section{INTRODUCTION}
Recent discoveries of electron-nematic phases, which break a certain point group symmetry of the system, have suggested that the superconducting pairing mechanism may be closely related to nematicity in some correlated electron systems, such as cuprates, iron-based compounds, heavy-fermions, doped-Bi$_2$Se$_3$, and magic-angle twisted-bilayer graphene (MA-TBG)\cite{
YCao_Nat2018_Ins,YCao_Nat2018_USC,
MYakowitz_Science2019,XLu_Nat2019,ALSharpe_Science2019,MSerlin_science2020,
YXie_Nature2019,RBistritzer_PNAS2011,NFYuan_PRB2018_MIT,
JKang_PRX2018_Wannier,MKoshino_PRX2018_Wannier,
HCPo_PRX2018_Origin,Zou_PRB2018_Wan,HCPo_PRB2019_Faithful}. Obviously, the relation between electron-nematic order and unconventional superconductivity is a pressing question in present condensed matter physics\cite{Fradkin_Ann2010,Fradkin_RMP2015,Fernandes_Ann2019,
CXu_PRL2018,Isobe_PRX2018,Venderbos_PRB2018,Kozii_PRB2019,
Chichinadze_PRB2020,YWang_PRB2021,Fernandes_PRL2021}.

In the case of MA-TBG, an electron-nematic state, which breaks the lattice $C_{3z}$ symmetry, has been detected by scanning tunneling microscopy\cite{YChoi_NatPhys2019,AKerelsky_Nat2019,YJiang_Nat2019} and transport measurements\cite{Cao_AX2020_NSC}. This $C_{3z}$-broken electron-nematic state, referred to as a three-state Potts nematic state, is of interest for its competition with nematic superconductivity\cite{Cao_AX2020_NSC} and for the mystery of the Landau level degeneracy\cite{Cao_AX2020_NSC,SLiu_AX2019,YHZhang_PRB2019} in different regions of its phase diagram\cite{AKerelsky_Nat2019,Cao_AX2020_NSC}. 
From a theoretical point of view\cite{Venderbos_PRB2018,YHZhang_PRB2019,Sboychakov_PRB2020,
Chichinadze_PRB2020,Fernandes_AdvSci2020,DParker_PRL2021,Brillaux_arxiv2020,Onari_AX2020}, it has been pointed out that unique properties of the moir\'e phonon, which reflects a non-rigid crystal\cite{Koshino_PRB2019,Ochoa_PRB2019}, assist a nematic phase transition\cite{Fernandes_AdvSci2020}, and the microscopic origin of this nematic state is attributed to the interference of the valley+spin fluctuation\cite{Onari_AX2020}.
Moreover, in the case of doped-Bi$_2$Se$_3$, which is a candidate material of nematic superconductors
\cite{Matano_NatPhys2016,Yonezawa_NatPhys2017,Pan_SciRep2016,Du_ScienceCP2017,Shen_npj2017,Asaba_PRX2017,Smylie_SciRep2018,Tao_PRX2018,Yonezawa_Cond2019}
, a three-state Potts nematic state has been reported\cite{Kuntsevich_NJP2018,YSun_PRL2019,Cho_Nat2020} above the superconducting transition temperature.
Although this seems to contradict the nematic superconductivity for which an order parameter is accompanied with a breaking of the lattice point group symmetry, it is pointed out that this nematic state is a vestige\cite{Fradkin_RMP2015} of the nematic superconductivity\cite{Hecker_npj2018,Fernandes_Ann2019,Cho_Nat2020} caused by the strong superconducting fluctuation. 
Besides the relationship between nematicity and superconductivity, it is also important to identify the critical behavior of electron-nematic states and to distinguish whether it is intrinsic (i.e. induced spontaneously) or extrinsic (i.e. due to trivial strains or the structural distortion).

Motivated by recent studies of the three-state Potts nematic state, we investigate the impact of critical nematic fluctuations on phonons, which in turn enables us to identify the nematic properties by ultrasound attenuation experiments. Despite a lot of research, the identification of such a three-state Potts nematic state and the clarification of whether it is induced spontaneously or from trivial strains are not an easy task. We analyze the influence of the nemato-elastic coupling on the low-energy properties of phonons by a phenomenological argument using a Ginzburg-Landau-Wilson (GL) action\cite{Altland_2010} 
and a model calculation based on the Hubbard model. It is shown that nematic fluctuations induce an isotropic divergence of the transverse sound attenuation coefficient, which is defined as the inverse of the phonon mean free path. 

The plan of this paper is as follows. In Sec. II, we present a phenomenological argument to see how the critical nematic fluctuation affects the properties of phonon. In Sec. III, we present a model calculation of nematicity, and we discuss a mean-field phase diagram. In Sec. IV, we give a brief discussion on the application of our results. Section V is devoted to a summary of the paper.
\section{PHENOMENOLOGICAL APPROACH}

In this section, we present a phenomenological theory to show how the ultrasound attenuation detects the critical nematic fluctuations. In the following subsection, we use models [see Eqs. (\ref{eq:GL_action}) and (\ref{eq:GL_coupling})] that agree with the pioneering work presented in Ref. \onlinecite{Fernandes_AdvSci2020}.
Because we consider how to capture the signature of the intrinsic nematic phase transition, our focus is different from Ref. \onlinecite{Fernandes_AdvSci2020}, where the nematicity affected by the static strain and acoustic phonons was discussed.

\subsection{GL action for nematic fluctuations}
First, we deal with the nematic phase transition phenomenologically.
In hexagonal lattices, such as MA-TBG and doped-Bi$_2$Se$_3$, the nematic order is described by a two-component order parameter $\bm{\Phi}=(\Phi_1,\Phi_2)$, which belongs to a two-dimensional representation of the point group D$_3$\cite{Onari_AX2020}, D$_6$\cite{Fernandes_AdvSci2020}, and D$_{3d}$\cite{Hecker_npj2018}, in the three-state Potts-model class. 
The GL action for the nematic fluctuation\cite{Fernandes_AdvSci2020} is given by
\begin{eqnarray}
S_{\rm{nem}}[\bm{\Phi}]&=&\int_x \Bigr[\frac{1}{2}r \Phi_+\Phi_- +\frac{1}{6}u_3 (\Phi_+^3+\Phi_-^3)+\frac{1}{4}u_4 (\Phi_+\Phi_-)^2\Bigl],\nonumber \\
\label{eq:GL_action}
\end{eqnarray}
where $x=(\bm{r},\tau)$, $\Phi_{\pm}=\Phi_{1}(x) \pm i\Phi_{2}(x)$, and GL coefficients $r, u_3, u_4$. 
$\bm{\Phi}$ is naturally parametrized as $\bm{\Phi}=\Phi(\cos{2\theta},\sin{2\theta})$, where the angle $\theta$ can be identified with the orientation of the nematic director $\hat{n}=(\cos{\theta},\sin{\theta})$ with angle $2\theta$ reflecting the invariance of $\pi$ rotation.
The cubic term reflects the hexagonal anisotropy and is expressed as 
\begin{eqnarray}
\frac{1}{6}u_3 (\Phi_+^3+\Phi_-^3)&=&\frac{1}{6}u_3 \Phi^3 \cos{6\theta},
\label{eq:GL_cubic}
\end{eqnarray}
which is minimized at $\theta=2n \pi/6=\{0, \pi/3, 2\pi/3\}$ for $u_3<0$ and $\theta=(2n+1) \pi/6=\{\pi/6, \pi/2, 5\pi/6\}$ for $u_3>0$. These solutions represent threefold degenerate nematic directors.

When we consider the Gaussian fluctuation region, the corresponding action for nematic fluctuation is given by
\begin{eqnarray}
S_{\rm{Gauss}}[\bm{\Phi}]&=&\int_q \bm{\Phi}_{q} \Bigl[ 
\hat{\chi}_{d}^{-1}(\bm{q},i\epsilon_m) \Bigr]\bm{\Phi}_{q}^*, 
\label{eq:gauss}
\end{eqnarray}
with $q=(\bm{q},i\epsilon_m)$, the boson Matsubara frequency $\epsilon_m$, and $\Phi^*_{iq}=\Phi_{i-q}$, because of $\Phi_i(x) \in \mathbb{R}$. Here,
\begin{eqnarray}
\hat{\chi}_{d}^{-1}(\bm{q},i\epsilon_m)=(r+\xi_0^2\bm{q}^2)\1+\hat{D}\Bigl(\frac{|\epsilon_m|}{\Gamma_{d}(\bm{q})}\Bigr),
\label{eq:nemati_propargator}
\end{eqnarray}
is the matrix of the $d$-wave density correlation function, 
where $r \propto T_{c0}-T$ measures the distance from the mean-field transition temperature $T_{c0}$, with the mean-field correlation length $\xi_0$ and the damping rate $\Gamma_{d}(\bm{q})$.
The Landau damping term $\hat{D}\Bigl(\frac{|\epsilon_m|}{\Gamma_{d}(\bm{q})}\Bigr)$ depends on the type of order parameter and the microscopic details of the system.

In the following subsection, we derive the functional form of $\hat{D}\Bigl(\frac{|\epsilon_m|}{\Gamma_{d}(\bm{q})}\Bigr)$ coming from the $C_{3}$-breaking bond-order [see Eq. (\ref{eq:gaussian_fluctuation})], which is an example of the three-state Potts nematic order.
Remarkably, we find that the $C_{3}$-breaking case has an isotropic angular dependence of the Landau damping, in sharp contrast to the strong angle dependence of the Landau damping in the case of the $C_4$-breaking bond-order\cite{Gallais_CRPhys2016,Paul_PRL2017}, which is an example of the Ising nematic order.

\subsection{Phenomenology of a $C_3$-breaking bond-order fluctuation}

According to the standard Hertz-Millis-Moriya description\cite{Hertz_PRB1976,Millis_PRB1993,SCR,Wolfle_RMP2007}, the dynamics of a ferroic order parameter which couples to an itinerant electron system is overdamped at low frequency. This is based on the simplest treatment of the critical order parameter fluctuation. 
On the other hand, the dynamics of electron-nematicity is more complicated\cite{Oganesyan_PRB2001,Wolfle_RMP2007,Fradkin_Ann2010}. 
For example, in isotropic Fermi liquids, the order parameter fluctuation of the $d$-wave Pomeranchuk instability is decomposed into a ballistic ($z=2$) transverse mode and an overdamped ($z=3$) longitudinal mode, where $z$ is a dynamical critical exponent. This nature leads to various intriguing properties unique to the nematic quantum critical point, such as an unusual non-Fermi-liquid behavior\cite{Garst_PRB2010,Yamase_PRB2011,SSLee_AnnRev2018} and the multiscale quantum criticality\cite{Zacharias_PRB2009}.
Moreover, in lattice systems with $C_4$-breaking bond-order fluctuation, the appearance of a ballistic mode and its effect on the critical properties have been discussed\cite{Gallais_CRPhys2016,Paul_PRL2017}.

Now we ask what happens for the dynamics of the nematic fluctuation for the $C_3$-breaking bond-order case, which is one of the microscopic origins of electron-nematicity (see Appendix \ref{appendix:nematicpolarization}).
For simplicity, we assume a circular Fermi surface around the $\Gamma$ point.
The interaction between the nematic fluctuation ($\Phi_{1\bm{q}},\Phi_{2\bm{q}}$) and the electrons ($c^{\dagger}_{\bm{k}},c_{\bm{k}}$) resulting from the Hubbard-Stratonovich transformation is given by 
\begin{eqnarray}
\mathcal{H}_{\rm coup}&\propto &\sum_{\bm{q},\bm{k}}
\Bigl[d_{1\bm{k}}\Phi_{1\bm{q}}+d_{2\bm{k}}\Phi_{2\bm{q}}\Bigr]
c^{\dagger}_{\bm{k}+\bm{q}/2}c_{\bm{k}-\bm{q}/2},
\end{eqnarray}
with form factors $d_{1\bm{k}} \sim (\hat{k}_x^2-\hat{k}_y^2)=\cos{2\theta_{\bm{k}}}$ and $d_{2\bm{k}}\sim 2(\hat{k}_x \hat{k}_y)=\sin{2\theta_{\bm{k}}}$. 
$\theta_{\bm{k}}$ represents the propagating direction of the wave vector $\bm{k}=|\bm{k}|(\hat{k}_x,\hat{k}_y)=|\bm{k}|(\cos{\theta_{\bm{k}}},\sin{\theta_{\bm{k}}})$.
It reflects a two-dimensional representation of a $C_3$ symmetric lattice, meaning that two waves, the $d_{x^2-y^2}$-wave and the $d_{xy}$-wave, cannot be treated separately.
The coupling term is expressed in terms of the relative angle between the wave vector and the nematic director $(\theta_{\bm{k}}-\theta)$ as follows, 
\begin{eqnarray}
\mathcal{H}_{\rm coup}&\propto & \sum_{\bm{q},\bm{k}}
\Phi_{\bm{q}}\cos{2(\theta_{\bm{k}}-\theta)}
c^{\dagger}_{\bm{k}+\bm{q}/2}c_{\bm{k}-\bm{q}/2}, 
\label{eq:electron_nematic_coupling}
\end{eqnarray}
where we have used $\Phi_{1\bm{q}}=\Phi_{\bm{q}}\cos 2\theta$, $\Phi_{2\bm{q}}=\Phi_{\bm{q}}\sin 2\theta$, and the coupling term vanishes at $\theta_{\bm{k}}-\theta=\pm \pi/4$.

The low-energy contribution of a nematic polarization matrix $\chi_{q}^{ij}=\sum_{k} d_{i\bm{k}}d_{j\bm{k}} G_{k}G_{k+q}$ with $i,j=1,2$ and an electron Green's function $G_{k}$ determines the dynamical properties of the nematic polarization $D_q^{ij}=\chi_q^{ij}-\chi_{\bm{q},0}^{ij}$. The $\bm{k}$-summation can be performed by linearizing the electronic dispersion, leading to
\begin{eqnarray}
D_{q}^{ij}&=&-ia \rho_{0}
\int_0^{2\pi}\frac{d\psi}{2\pi}
\frac{d_{i\bm{k}}d_{j\bm{k}} }{ia-\cos{\psi}}, 
\end{eqnarray}
with $\psi=(\theta_{\bm{k}}-\theta_{\bm{q}})$, $a=\frac{\epsilon_m}{v_{\rm F}|\bm{q}|}$, the density of states at the Fermi level $\rho_{0}$, the Fermi velocity $v_{\rm F}$, and the boson Matsubara frequency $\epsilon_m$.
After evaluating the above integration, the dynamical part of the nematic polarization matrix in the static region ($|\epsilon_m|\ll v_{\rm F}|\bm{q}|$) is,
\begin{eqnarray}
\hat{D}_{q}
&=&
-\rho_{0}\frac{|a|}{2}\1
-\rho_{0}\Bigl[ \frac{|a|}{2}-2a^2\Bigr]
\left(
\begin{array}{cc}
\cos{4\theta_{\bm{q}}}& \sin{4\theta_{\bm{q}}}\\
\sin{4\theta_{\bm{q}}} & -\cos{4\theta_{\bm{q}}}
\end{array}
\right).\nonumber \\
\end{eqnarray}
At first glance, this would seemingly break the $C_3$-symmetry, but later calculations show that the $C_3$-rotation symmetry is preserved 
when the angle of the nematic directors is taken into account.
Next, we express $\hat{D}_q$ in terms of the angle $\theta$ of nematic directors.
Thus the Gaussian action including the above discussion is rewritten as
\begin{eqnarray}
S_{\rm{Gauss}}[\bm{\Phi}]&=&\int_q
\bm{\Phi}^T_{q}
\Bigl[
\bigl(r+\xi_0^2\bm{q}^2\bigr)\1+\hat{D}_q
\Bigr]\bm{\Phi}^*_{q}, \\ 
\bm{\Phi}^T_{q}\hat{D}_{q}\bm{\Phi}^*_{q}
&=&
-\Phi(q)
\rho_{0}\Bigl[
\frac{|\epsilon_m|}{v_{\rm F}|\bm{q}|}\cos^2{(2\theta_{\bm{q}}-2\theta)}\nonumber \\
&&-2\frac{|\epsilon_m|^2}{(v_{\rm F}|\bm{q}|)^2} \cos{(4\theta_{\bm{q}}-4\theta)}\Bigr]\Phi^*(q).
\label{eq:dynamics}
\end{eqnarray}
The orientation of the nematic directors is restricted to three directions by the cubic term as follows: 
$\theta=\{0, 2\pi/3, 4\pi/3\} $ for $u_3<0$ and $\theta=\{-\pi/6, \pi/2, 7\pi/6\}$ for $u_3>0$. 
Precisely speaking, the damping term preserves this $\mathbb{Z}_3$ symmetry in a disordered state, thus we need to treat three angles $\theta$ equivalently; 
$\cos^2{(2\theta_{\bm{q}}-2\theta)} \rightarrow \frac{1}{3}\Bigl[ \cos^2{(2\theta_{\bm{q}})}+\cos^2{(2\theta_{\bm{q}}-\frac{2\pi}{3})}+\cos^2{(2\theta_{\bm{q}}-\frac{4\pi}{3})}\Bigr]=\frac{1}{2}$ for $u_3<0$. 
Eventually, we arrive at the following action with the single component scalar field $\Phi$:
\begin{eqnarray}
S_{\rm{Gauss}}[\Phi]&=&\int_q \Phi(q) \Bigl[ \chi^{-1}_d(q)\Bigr]\Phi^*(q),
\label{eq:gaussian_fluctuation}\\
\chi^{-1}_d(q)&=&r+\xi_0^2\bm{q}^2+
\frac{|\epsilon_m|}{\Gamma_d(\bm{q})},
\end{eqnarray}
with $\bm{\Phi}=\Phi(\cos{2\theta},\sin{2\theta})$ and the damping rate $\Gamma^{-1}_d(\bm{q}) =\frac{\rho_{0}}{2v_{\rm F}}|\bm{q}|^{-1}$. 
We conclude that the $C_3$-breaking bond-order fluctuation leads to an isotropic angular dependence of the Landau damping.

The above results are quite contrasted to the Ising nematic case where the nematic director is forced to be $\theta=\{0,\pi/2 \}$ for the $d_{x^2-y^2}$-wave. In that case, the term $D\Bigl(\frac{|\epsilon_m|}{\Gamma_{d}(\bm{q})}\Bigr)$ in Eq. (\ref{eq:nemati_propargator}) is expressed as the following anisotropic form\cite{Gallais_CRPhys2016,Paul_PRL2017}: 
$\Bigl[
\frac{|\epsilon_m|}{v_{\rm F}|\bm{q}|}\cos^2{2\theta_{\bm{q}}}-2\frac{|\epsilon_m|^2}{(v_{\rm F}|\bm{q}|)^2} \cos{4\theta_{\bm{q}}}\Bigr]$, which leads to the angle-dependent dynamics of nematic fluctuation. It is possible to understand from the coupling term in Eq. (\ref{eq:electron_nematic_coupling}) what is responsible for these differences between the three-state Potts nematicity and the Ising nematicity, as follows.
The dynamics of nematic fluctuation is damped due to particle-hole pair excitations close the Fermi surface, which is a source of the Landau damping.
It requires electrons to scatter along the Fermi surface. 
One of the unique properties of bond-orders is the presence of the nodal structure in the form factor\cite{Gallais_CRPhys2016,Paul_PRL2017}. This implies that a particle-hole pair creation is prohibited at certain directions, leading to a large anisotropy in physical quantities.
For example, in the case of the Ising nematicity, the nematic director is forced to be $\theta=\{0,\pi/2 \}$ for $d_{x^2-y^2}$-waves, so that the coupling term vanishes at $\theta_{\bm{k}}=\pm \pi/4$ in Eq. (\ref{eq:electron_nematic_coupling}).
On the contrary, the three-state Potts nematic case of our interest does not have such a specific direction of vanishing coupling because nematic directors are not orthogonal to each other, as we have discussed in this subsection.

\subsection{Probing the nematicity through acoustic phonons}

In addition to the angle dependence of the Landau damping $\hat{D}\Bigl(\frac{|\epsilon_m|}{\Gamma_{d}(\bm{q})}\Bigr)$ in Eq. (\ref{eq:nemati_propargator}), there is a unique character in the nematic order, i.e.,  
the nematic order parameter couples linearly to acoustic phonon modes\cite{HYKee_PRB2003_sound,Adachi_PRB2009,Karahasanovic_PRB2016,Paul_PRL2017,Hecker_npj2018,Carvalho_PRB2019,Fernandes_AdvSci2020}. 
This is essentially different from the cases of other ferroic orders, e.g., ferromagnetism or superconductivity, whose order parameters only couple to the totally symmetric mode of a phonon in quadratic order.
Because of this specific form of coupling, the unique properties are reflected in the transverse acoustic phonon.
As a result, through linear nemato-elastic coupling, phonon modes affect the thermodynamic and transport properties near the nematic critical point.

Despite a lot of research, an identification of the electron-nematic phase transition and clarifying whether it is induced spontaneously or from trivial strains is not an easy task. The ultrasound attenuation of acoustic phonons is one of the good techniques of identifying the electron-nematic phase transition and its critical behavior.
It is also pointed out that the selection rules of ultrasound attenuation coefficients can determine the Ising nematic phase transition\cite{Adachi_PRB2009}.
In this section, we focus on the impact of nemato-elastic coupling on acoustic phonons.

First we consider the dynamical properties of two acoustic phonon modes, a transverse ($T$) and a longitudinal ($L$) one, with sound velocity $v_{T(L)}$. The displacement field $\bm{u}$ is decomposed into two modes $\bm{u}_{\mu=T,L}=\tilde{u}_{\mu}\hat{\bm{e}}_{\mu}$ with $\hat{\bm{e}}_{T}=(-\sin{\theta_{\bm{q}}}, \cos{\theta_{\bm{q}}})$, $\hat{\bm{e}}_{L}=(\cos{\theta_{\bm{q}}}, \sin{\theta_{\bm{q}}})$, and $\theta_{\bm{q}}=\tan{^{-1}(q_y/q_x)}$. 
The elastic action for two acoustic phonon modes reads\cite{Altland_2010},
\begin{eqnarray}
S_{\rm{ph}}[\bm{u}]&=&\frac{\rho}{2}\sum_{\mu=T, L} \int_q \tilde{u}_{\mu}(q)
K_{\mu}(q)\tilde{u}^*_{\mu}(q),\\
K_{\mu}(q)&=&K^{(0)}_{\mu}(q)-\delta K_{\mu}(q),
\label{eq:phonon_full}
\end{eqnarray}
with $q=(\bm{q},i\epsilon_m)$, the full (bare) inverse propagator $K_{\mu}$($K_{\mu}^{(0)}$), the phonon self-energy $\delta K_{\mu}$, the boson Matsubara frequency $\epsilon_m=2\pi Tm$, and the mass density $\rho$. The bare inverse propagator has the form $K_{\mu}^{(0)}=\epsilon_m^2+v^{2}_{\mu}\bm{q}^2$. The sound attenuation coefficient\cite{AGD} $\alpha_{\mu}$ is defined as the inverse of the phonon mean-free path, as follows:
\begin{eqnarray}
\alpha_{\mu}(\bm{q})&=&-\lim_{\omega \to 0}\frac{1}{v_{\mu}\omega}{\rm Im}K^{R}_{\mu}(\bm{q},\omega),
\end{eqnarray}
where $K^{R}_{\mu}(\bm{q},\omega)$ is the retarded function of the full inverse propagator.

In general, the lowest order of the symmetry-allowed nemato-elastic coupling\cite{Fernandes_AdvSci2020,Hecker_npj2018} in the free energy is
\begin{eqnarray}
F_{\rm{nem-ph}}[\bm{\Phi},\bm{u}]&=&-\kappa \int_{\bm{r}} \Bigl[(\epsilon_{xx}-\epsilon_{yy})\Phi_1+2\epsilon_{xy}\Phi_2 \Bigr],
\label{eq:GL_coupling}
\end{eqnarray}
with the coupling constant $\kappa$ and the strain tensor $\epsilon_{ij}=\frac{1}{2}(\partial_i u_j+\partial_j u_i)$. 
Considering the nemato-elastic coupling $S_{\rm{nem-ph}}=\int_{\tau}F_{\rm{nem-ph}}$, we calculate the effective action for phonons coupled with nematic fluctuation.
In terms of $\tilde{u}_{L}(q)$ and  $\tilde{u}_{T}(q)$, the nemato-elastic action reads
\begin{eqnarray}
S_{\rm{nem-ph}}[\Phi,\bm{u}]&=&-\kappa \int_q 
\left(
\begin{array}{cc}
\tilde{u}_{L}(q) & \tilde{u}_{T}(q)
\end{array}
\right)
\nonumber \\
&\times &i|\bm{q}| 
\left(
\begin{array}{c}
\cos{(2\theta_{\bm{q}}-2\theta)} \\ 
-\sin{(2\theta_{\bm{q}}-2\theta)} 
\end{array}
\right)
\Phi^*(q), \nonumber \\
\end{eqnarray}
with the angle of nematic director $\theta$ (see Appendix \ref{appendix:nematoelasticcoupling}).
Treating the three angles equally does not show anisotropy, with a similar argument as before, and thus we obtain the following form
\begin{eqnarray}
S_{\rm{nem-ph}}[\Phi,\bm{u}]&=&-\kappa  \int_q  i\frac{|\bm{q}|}{2}
\Bigl[ \tilde{u}_{L}(q) -\tilde{u}_{T}(q)\Bigr]\Phi^*(q). \nonumber \\
\label{eq:nemato_elastic_coupling}
\end{eqnarray}
Therefore we conclude that the nemato-elastic coupling has an isotropic angular dependence.
After integrating out the nematic order parameter field in the total action $S_{\rm tot}=S_{\rm{Gauss}}[\Phi]+S_{\rm{ph}}[\bm{u}]+S_{\rm{nem-ph}}[\Phi,\bm{u}]$, an additional contribution to the phonon Green's function in Eq. (\ref{eq:phonon_full}) is 
\begin{eqnarray}
\delta K_{\mu}(q)=\frac{\kappa^2\bm{q}^2}{2\rho}\chi_d(q).
\end{eqnarray}
Indeed, up to the leading order correction, we can confirm that the self-energy has no anisotropy.

As a consequence, we obtain the full inverse propagator for phonons in Eq. (\ref{eq:phonon_full}), 
which gives rise to the renormalization of sound velocities as,
\begin{eqnarray}
v^*_{\mu}&=&v_{\mu}\sqrt{1-\frac{{\rm Re}\delta K^R_{\mu \mu}(\bm{q},\omega \rightarrow 0)}{v^2_{\mu}q^2}}, \nonumber \\
&=&v_{\mu}\sqrt{1-\frac{\kappa^2}{2 v^2_{\mu}\rho}{\rm Re}\chi^R_{d}(\bm{q},\omega \rightarrow 0)}.
\end{eqnarray}
Note that a sound velocity renormalization implies a lattice softening.
They are tied together in the following equation $v_{\mu}=\sqrt{c_{\mu}/\rho}$, where the corresponding elastic constants are $c_{\mu}$.
In the same way, sound attenuation coefficients are 
\begin{eqnarray}
\alpha_{\mu}(\bm{q})&=&-\lim_{\omega \to 0}\frac{1}{v^*_{\mu}\omega}{\rm Im}\delta K^{R}_{\mu \mu}(\bm{q},\omega), \nonumber \\
&=&\lim_{\omega \to 0}\frac{\kappa^2\bm{q}^2}{2\rho v^*_{\mu}\omega}{\rm Im}\chi_d^R(\bm{q},\omega \rightarrow 0),
\label{eq:sound_attenuation}\\
&\sim&  \frac{\kappa^2 }{2\rho v^*_{\mu}}
\frac{1}{r^2 }\frac{|\bm{q}|}{\gamma_d},
\end{eqnarray}
with $\gamma_d=\frac{2v_{\rm F}}{\rho_0}$.
Thus $\alpha_{\nu}(\bm{q})\propto r^{-2}$. The symmetry-allowed coupling term leads to the isotropic divergence of transverse (longitudinal) sound attenuation $\alpha_{T(L)} \propto (T_c-T)^{-2}$ and an isotropic lattice softening.

In addition to the above equation, there is another relevant term\cite{Paulson_PLA1968,Adachi_PRB2009} which is induced by the deformation potential, 
\begin{eqnarray}
F^{\prime}_{\rm{nem-ph}}[\bm{\Phi},\bm{u}]
&=&\kappa^{\prime}\sum_{\bm{q},\bm{q}^{\prime}}\Phi^{*}_a(\bm{q}+\bm{q}^{\prime})
\Phi_a(\bm{q}^{\prime})[i|\bm{q}|u_L(\bm{q})],\nonumber \\
\end{eqnarray}
where the longitudinal sound modes couple to the quadratic term of nematic fields. It originates from a change in volume due to the effective nematic-nematic interaction. This term also leads to the divergent contribution to the longitudinal sound attenuation $\alpha_L\propto (T_c-T)^{-2}$. Note that the latter term is essentially the same as in weak ferromagnetism\cite{Paulson_PLA1968} for sound attenuation near the ferromagnetic transition in metals. 
\footnote{Paulson and Schrieffer considered that the deformation potential of the effective exchange interaction $J$ gives an interaction between a phonon and the electron spins of the form
\begin{eqnarray}
H_{\rm{el-spin}}[\bm{M},\bm{u}]&=&-J \int_V  \bm{\nabla}\cdot\bm{u}(\bm{r}) \, \bm{M}(\bm{r})\cdot\bm{M}(\bm{r}),
\end{eqnarray}
where $\bm{M}(\bm{r})=\psi^{\dagger}(\bm{r})\bm{\sigma}\psi(\bm{r})$ is the electron spin density.
They pointed out that $\alpha_L\propto \omega^2(T-T_c)^{-2\gamma}$ above $T_c$.
}

Finally, we comment on the comparison with the Ising nematic case.
In the case of the Ising nematicity, the nematic director is forced to be $\theta=\{0,\pi/2 \}$. 
Even if we treat the two angles equally, the anisotropy of the nemato-elastic coupling remains.
As pointed out in previous studies, this leads to the angle dependent damping properties of acoustic phonons\cite{Adachi_PRB2009} or the mass term anisotropy of the Ising nematic fluctuation\cite{Paul_PRL2017}.

We conclude that the following unique properties illustrate the three-state Potts nematic order: (i) the nematic fluctuation affects the transverse acoustic phonon, (ii) the ultrasound attenuation coefficients show an isotropic divergence which is proportional to the momentum $|\bm{q}|$ and (iii) the sound velocity renormalization also shows an isotropic angle dependence.
We therefore propose to detect the three-state Potts nematic order by measuring the isotropic divergence of the transverse sound attenuation coefficient and the isotropic sound velocity renormalization.
The isotropic nature of these properties is in contrast to the Ising nematic case where such quantities are anisotropic and subject to selection rules\cite{Adachi_PRB2009}.
Note that the vanishing anisotropy of the acoustic phonon velocity is consistent with Cowley's classification\cite{Cowley_PRB1976}.

\section{MODEL CALUCULATION}
Now we move to the model calculation of the nematic phase originating from a bond order on the honeycomb lattice. 
In this section, we use the mean-field approximation by taking into account the higher order terms up to the sixth order coefficients in Eq. (\ref{eq:GL_action}).
Since the critical properties near the phase transition are evaluated in the mean-field approximation, the power of divergence may be changed in the presence of strong fluctuations, but the stability and the extent of the ordered phase are expected to remain qualitatively unchanged even with the inclusion of such effects of the mode coupling.

\subsection{Model and method}
In a TBG, a slight mismatch in the lattice periods of two graphene layers gives rise to a long-period moir\'e interference pattern.
The regions that locally appear to be AB-stacked bilayer grahene and BA-stacked bilayer graphene form the emergent honeycomb lattice\cite{JKang_PRX2018_Wannier,MKoshino_PRX2018_Wannier,Zou_PRB2018_Wan}. 
Now we focus on the electron-nematic phase transition near the van-Hove (VH) filling where the nematicity can be seen in the experiment, as claimed in a previous theoretical study\cite{Onari_AX2020}.
These authors showed that the $C_3$-breaking bond ordered state is stabilized near VH filling by using the so-called DW equation method beyond our mean-field description. Based on this work\cite{Onari_AX2020}, we restrict ourselves to the $d$-wave forward scattering channel of electron-electron interactions only.
The forward-scattering model\cite{Yamase_PRB2007,HYKee_PRB2003,Khavkine_PRB2004,Yamase_PRB2005,Valenzuela_NJP2008} derived from an extended Hubbard model on the emergent honeycomb lattice (see Appendix \ref{appendix:interaction}) reads
\begin{eqnarray}
H&=&\frac{1}{N}\sum_{\bm{k}\xi\sigma}
\left(
\begin{array}{cc}
c_{\bm{k}\xi\sigma}^{AB\dagger} & c_{\bm{k}\xi\sigma}^{BA\dagger}  
\end{array}
\right)
\hat{\mathcal{H}}^{ \xi}_{\bm{k}\sigma}
\left(
\begin{array}{c}
c_{\bm{k}\xi\sigma}^{AB} \\
c_{\bm{k}\xi\sigma}^{BA}  
\end{array}
\right) +H_{\rm{int}}+H_{\rm{imp}}
,\label{eq:micro_hamiltonian}
\nonumber \\ \\
H_{\rm{int}}&=&-g \sum_{\xi \sigma}\sum_{i=1,2}\sum_{\bm{q}}
\Bigl(n^{AB \xi\sigma}_{E_i}(\bm{q})n^{BA\xi\sigma}_{E_i}(-\bm{q})\Bigr),\nonumber \\
\label{eq:forward_scattering}
\end{eqnarray}
with creation and annihilation operators $c_{\bm{k}\xi\sigma}^{\alpha \dagger},c_{\bm{k}\xi\sigma}^{\alpha}$, the spin index $\sigma$, the sublattice index $\alpha \in \{ \rm{AB}, \rm{BA}\}$, the valley index $\xi \in \{+,- \}$, and the coupling constant $g=\frac{2V_{\rm{NN}}}{3}$ ($V_{\rm NN}$ is the nearest-neighbor repulsive interaction).
The above forward-scattering interaction or the long-range interaction comes from the three-peak structure of Wannier orbitals in MA-TBG\cite{JKang_PRX2018_Wannier,MKoshino_PRX2018_Wannier,Zou_PRB2018_Wan}.
Here $\mathcal{H}^{\xi}_{\bm{k}\sigma}$ is a $2\times 2$ Hamiltonian for each valley $\xi$ and spin $\sigma$.

Since our mean-field analysis aims at showing the critical properties of the nematic fluctuation and order of the metallic phase, we use a reduced tight-binding model with only the nearest-neighbor hopping term on the honeycomb lattice and deal with all spin and valley degrees of freedom on an equal footing. Although the band structure is somewhat different from the Bistritzer-MacDonald model and other tight-binding models\cite{RBistritzer_PNAS2011,NFYuan_PRB2018_MIT,
JKang_PRX2018_Wannier,MKoshino_PRX2018_Wannier,
HCPo_PRX2018_Origin,Zou_PRB2018_Wan,HCPo_PRB2019_Faithful}, our simple model captures the essential properties around VH filling, including the correlated insulating phase near half-filling. Imposing valley-$U$(1) symmetry, we introduce two orbitals which do not hybridize with each other. Each valley for $\xi=\pm$ is independent in the non-interacting Hamiltonian. Although the Coulomb interaction term may have both contributions from the intra-valley and the inter-valley interaction, the obtained form factor from the DW-equation methods\cite{Onari_AX2020} has no inter-valley component. In our mean-field calculation, we analyze all spin and valley degrees of freedom on an equal footing in the following section. As is known, in order to reproduce the correlated insulating phase near VH filling, which is not expected in ordinary single layer graphene\cite{Black_Schaffer_JPhys2014}, valley degrees of freedom are needed. In this paper, we focus on the nematic metallic phase with the $C_3$-breaking Fermi surface, in line with the transport measurement in Ref. \onlinecite{Cao_AX2020_NSC}.
The interaction term is shown in Appendix \ref{appendix:interaction} in terms of the $d_i$-wave density operator $n^{AB\xi \sigma}_{E_i}(\bm{q})=\frac{1}{N}\sum_{\bm{k}}E^{i*}_{\bm{k}}c^{AB\xi \dagger}_{\bm{k}+\bm{q}/2\sigma}c_{\bm{k}-\bm{q}/2\sigma}^{BA\xi}$, where $E^{i*}_{\bm{k}}$ are form factors in a two-dimensional $E$ representation.

The third term $H_{\rm{imp}}$ in Eq. (\ref{eq:micro_hamiltonian}) represents the spin-independent short-range isotropic impurity scattering,
\begin{eqnarray}
H_{\rm{imp}}&=&\sum_{\xi \sigma \alpha i}u^{{\rm imp}\xi \sigma \alpha }_{i} n^{\xi \sigma \alpha }_{i},
\end{eqnarray}
where the random impurity potential $u^{{\rm imp}}$ obeys the Gaussian ensemble $\langle u^{{\rm imp}}_i\rangle=0, \langle u^{{\rm imp}}_i u^{{\rm imp}}_j\rangle=n_{\rm imp}|u|^2\delta_{i,j}$ with $n_{\rm imp}$ and $u$ being the impurity concentration and the strength of the impurity potential. We resort to the Born approximation, which results in the impurity-averaged self-energy
\begin{eqnarray}
\hat{\Sigma}_{\rm{imp}}^{\xi \sigma \alpha}(i\omega_n)&=& n_i|u|^2\frac{T}{N}\sum_{\bm{k}}\hat{G}(\bm{k},i\omega_n),\nonumber \\
&=&i \Gamma {\rm sign}(\omega_n)\1,
\label{eq:born_selfenergy}
\end{eqnarray}
where $i\omega_n$ is the Matsubara frequency and $\Gamma$ is the strength of the impurity scattering. 
In this calculation, we use Eq. (\ref{eq:born_selfenergy}) or its retarded representation.
With this approximation, the impurity-averaged Green's function is solved as $\hat{G}^{-1}(k)=\hat{G}^{-1}_{0}(k)-\hat{\Sigma}_{\rm{imp}}(k)$.

Next, we introduce the two-component nematic order parameter field $\bm{\Phi}(q)$, with $\bm{\Phi}=(\Phi_1,\Phi_2)$.
After integrating out the electron degrees of freedom, we have an effective action (see Appendix \ref{appendix:action}), 
\begin{eqnarray}
S_{\rm eff}[\bm{\Phi}]
&=&g^{-1}\sum_{i\xi \sigma}\int_q\Phi_i(-q)\Phi_i(q) -{\rm Tr}{\rm ln}\Bigl[  \hat{M}_{k+\frac{q}{2}, k-\frac{q}{2}}^{\xi \sigma}\Bigr] , \nonumber \\ 
\label{eq:action}\\
\hat{M}_{k+\frac{q}{2}, k-\frac{q}{2}}^{\xi \sigma}&=&
\bigl(-i\omega_n \1+\hat{\mathcal{H}}^{\xi}_{\bm{k}\sigma}\bigr)\delta_{k+\frac{q}{2},k-\frac{q}{2}}\nonumber \\&&
-\frac{\Phi_i(-q)}{\sqrt{\beta N}}
\left(
\begin{array}{cc}
0 & E^{i*}_{\bm{k}}\\
 E^{i}_{\bm{k}}& 0
\end{array}
\right),
\end{eqnarray}
with $q=(\bm{q},i\omega_n)$, $k=(\bm{k},i\omega_m)$ and the form factor $E_{\bm{k}}^i$ in Appendix \ref{appendix:interaction}, where we have neglected any loop-current order and only considered the $(d_{x^2-y^2},d_{xy})$-wave components for simplicity.

In terms of the order parameter field $\bm{\Phi}$, the partition function is expressed in a functional integral form, $Z=Z_0\int \mathcal{D}\bm{\Phi}e^{-S_{\rm{eff}}[\bm{\Phi}]}$ and the Landau free energy is given by $\exp{(-F/T)}=\int \mathcal{D}\bm{\Phi} e^{-S_{\rm eff}[\bm{\Phi}]}$, where the GL action up to the sixth order terms reads, 
\begin{eqnarray}
F_{\rm{nem}}[\bm{\Phi}_0]&=&
\frac{1}{2}r  \Phi_+\Phi_-
+ \frac{1}{6}u_3   (\Phi_+^3+\Phi_-^3)+ \frac{1}{4}u_4 (\Phi_+\Phi_-)^2\nonumber \\
&+&\frac{1}{10}u_5 (\Phi_+^4\Phi_-+ \Phi_+\Phi_-^4)+\frac{1}{6}u_6 \Phi_+^3\Phi_-^3,\nonumber \\
\label{eq:GLW_action_sixth}
\end{eqnarray}
with $\Phi_{\pm}=\Phi_{1}(0)\pm i\Phi_2(0)$, the uniform ($\bm{q}=\bm{0}$) and static ($i\omega_n=0$) component $\bm{\Phi}_0=\bm{\Phi}(q=0)$, and coefficients $u_n$ and $r$ defined in Appendix \ref{appendix:GLexpansion}.

To calculate the sound attenuation coefficients and the sound wave renormalization, we derive an electron-acoustic phonon coupling for arbitrary filling of the honeycomb lattice. The electron-phonon coupling arises from the lattice modulation by phonons, which leads to a change in the nearest neighbor hopping $t$, the so-called bond-length change\cite{Suzuura_PRB2002,CastroNeto_RMP2009,Vozmediano_PhysRep2010}. The detailed derivation is summarized in Appendix \ref{appendix:electron-phonon}.
The dominant contribution to the phonon self-energy in Eq. (\ref{eq:phonon_full}) is given by the bubble diagrams with electron-phonon vertices, 
\begin{eqnarray}
\delta K_{\mu, \rm{el-ph}}(q)&=&-\frac{g^2_{\rm ph}}{2\rho}\int_q {\rm tr} \Bigl[ \hat{G}_{k+q/2}\hat{w}^{\mu}_{k,q}\hat{G}_{k-q/2}\hat{w}^{\mu}_{k,-q}\Bigr], \nonumber \\
\label{eq:electron_phonon}
\end{eqnarray}
with 
\begin{eqnarray}
\hat{w}_{k,q}^{\mu}&=&
-\frac{g_{\rm ph}}{\sqrt{\beta N}}
\left(
\begin{array}{cc}
0 & \Delta \bm{E}^{*}_{\bm{k},\bm{q}}\cdot \hat{\bm{e}}_{\mu}(-q)\\
\Delta \bm{E}_{\bm{k},\bm{q}}\cdot \hat{\bm{e}}_{\mu}(-q)& 0
\end{array}
\right),\nonumber \\
\Delta \bm{E}_{\bm{k},\bm{q}}&=&
\left(
\begin{array}{c}
-\frac{1}{2} \\-\frac{\sqrt{3}}{2}
\end{array}
\right)e^{i\bm{k}\cdot \bm{a}_1}(i\bm{q}\cdot \bm{a}_1)+
\left(
\begin{array}{c}
-\frac{1}{2} \\ \frac{\sqrt{3}}{2}
\end{array}
\right)e^{i\bm{k}\cdot \bm{a}_2}(i\bm{q}\cdot \bm{a}_2), \nonumber \\  
\end{eqnarray}
where $\hat{\bm{e}}_{T}=(-\sin{\theta_{\bm{q}}}, \cos{\theta_{\bm{q}}})$ and $\hat{\bm{e}}_{L}=(\cos{\theta_{\bm{q}}}, \sin{\theta_{\bm{q}}})$ with $\theta_{\bm{q}}=\tan{^{-1}(q_y/q_x)}$.

\subsection{Mean-field phase diagram\label{section:mean-field}}

\begin{figure}
\includegraphics[width=7cm]{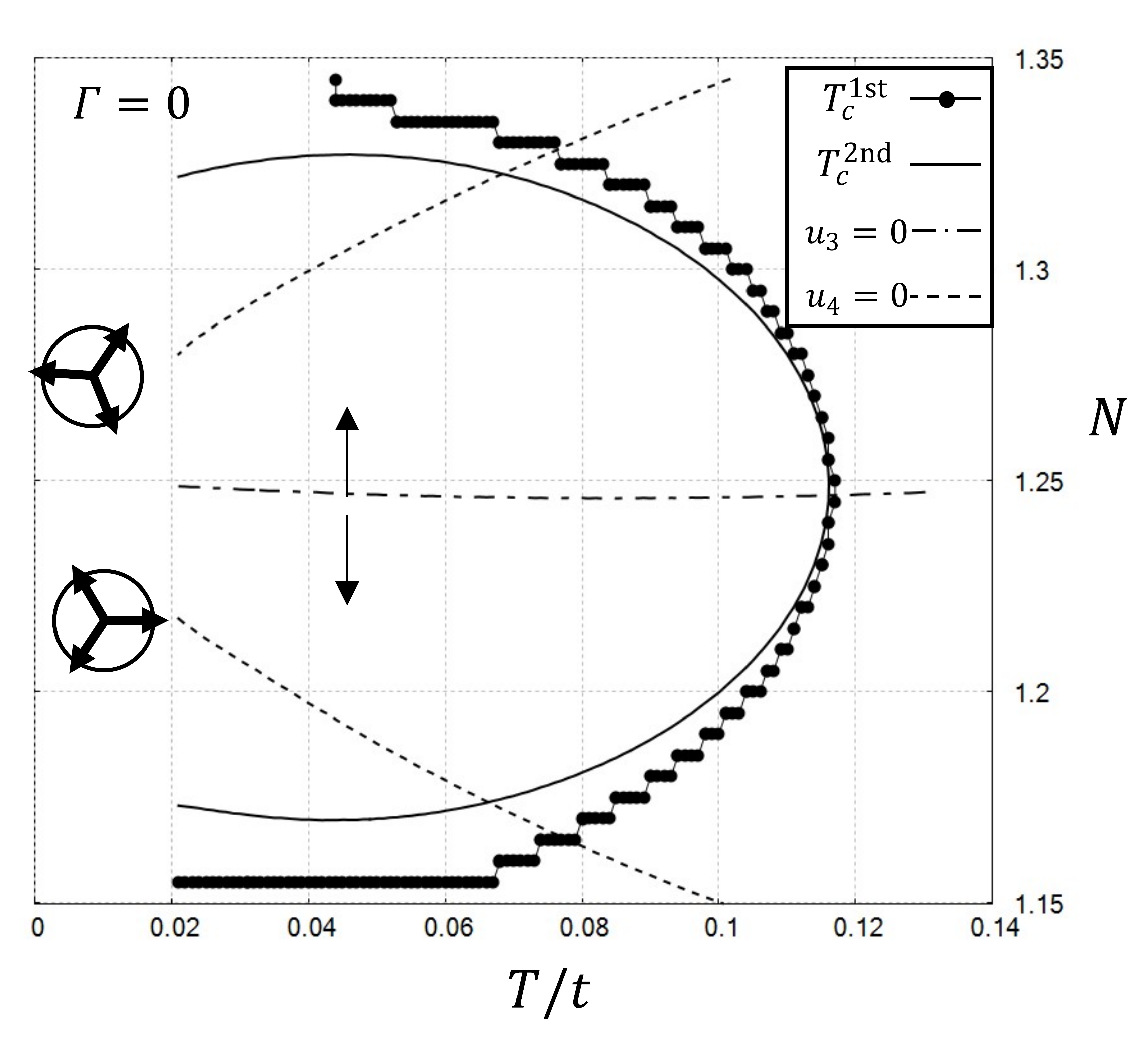}
\caption{
Phase diagram of a nematic bond-ordered state. 
We use $V_{\rm NN}/t=4.5$.
$T_c^{1{\rm st}}$ ($T_c^{2{\rm nd}}$) is the first (second) order phase transition point, and $u_3=0$ ($u_4=0$) is zeros of $u_3$ ($u_4$).
The three arrowheads surrounded by the circle represent the set of the orientation of the nematic director.
The set of the orientation changes on the zeros of $u_3$.
We calculate this by using a square mesh of $500\times 500$ in the Brillouin zone. 
The phase transition line $T_c^{1{\rm st}}$ is defined by $F_{\rm{nem}}[\Phi_1, \Phi_2]=0$ and $\partial F_{\rm{nem}}[\Phi_1, \Phi_2]/\partial \Phi_i =0$ with $i=1,2$.
}
\label{fig:phase}
\end{figure}

\begin{figure}
\includegraphics[width=9cm]{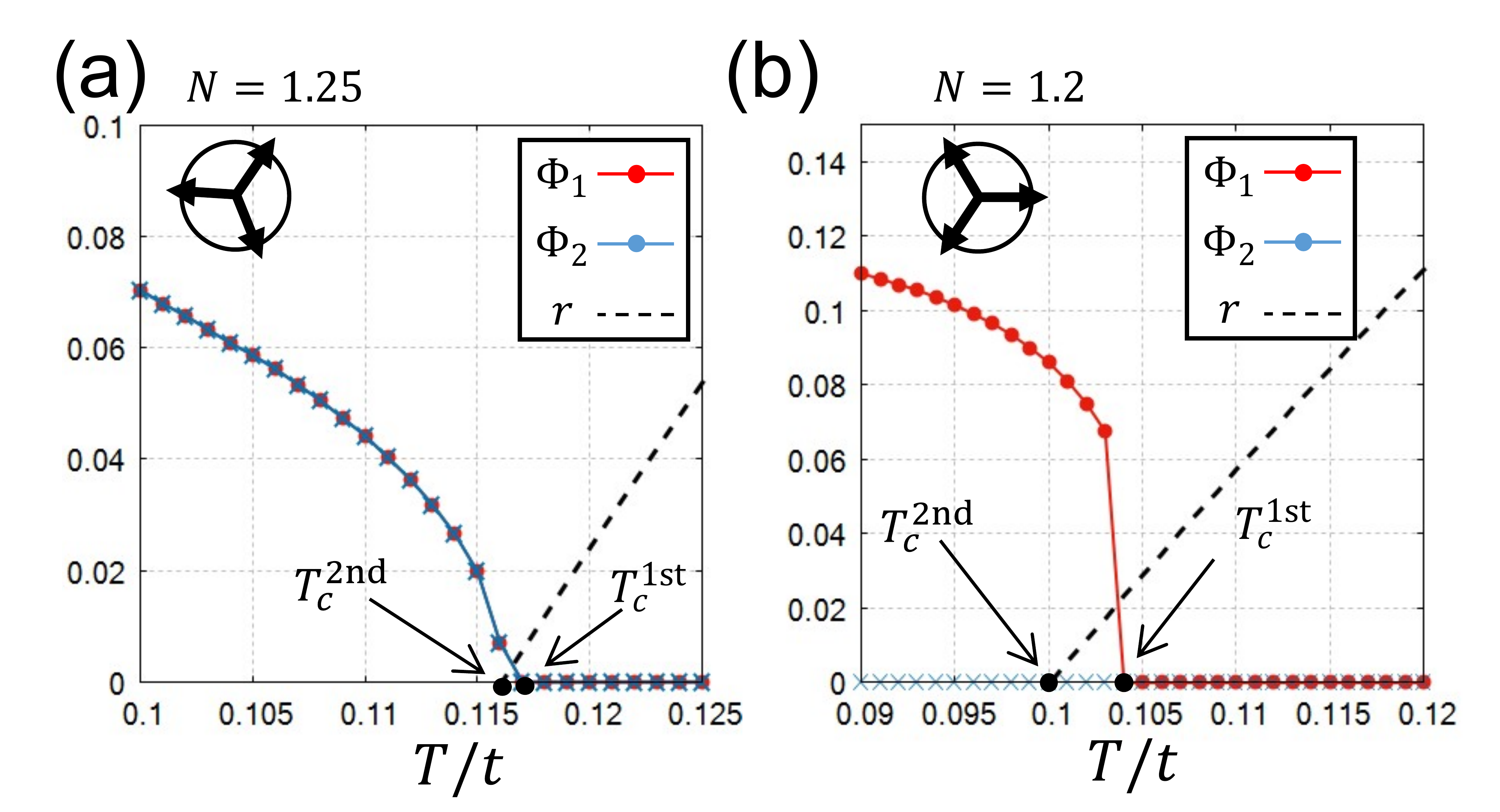}
\caption{
Nematic order parameters $(\Phi_1, \Phi_2)$ and $r \propto T_{c0}-T$ measures the distance from the mean-field transition temperature $T_{c}^{\rm 2nd}$.
The three arrowheads surrounded by the circle represent the set of the orientation of the nematic director.
$T_c^{1{\rm st}}$ ($T_c^{2{\rm nd}}$) is the first (second) order phase transition point.
(a) $N=1.25$ which is very close to the VHs. 
(b) $N=1.2$ which clearly shows the first order phase transition. 
}
\label{fig:nem}
\end{figure}

Now we determine Landau free energy coefficients up to the sixth order $(r, u_2,\cdots, u_6)$ numerically
\footnote{In this calculation, we use the band representation of the non-interacting Green's function.
The ($\alpha, \beta$)-component is expressed as 
$\Bigl[\mathcal{G}_{k}\Bigr]_{\alpha \beta}=\sum_{\gamma}\Bigl[U_{\bm{k}}\Bigr]_{\alpha \gamma}\Bigl[U_{\bm{k}}^{\dagger}
\Bigr]_{\gamma \beta}\it{g}^{\gamma}_{\bm{k},i\omega_m}$, where the non-interacting Hamiltonian $\mathcal{H}_{\bm{k}}$ and the unitary matrix $U_{{\bm{k}}}$ as follows:
$\mathcal{H}_{\bm{k}}U_{{\bm{k}}}=U_{{\bm{k}}}D_{\bm{k}}$, a diagonal matrix $D_{\bm{k}}={\rm diag}\{\epsilon^{\gamma}_{\bm{k}}\}$, $\gamma$-th component of eigenvalue $\epsilon^{\gamma}_{\bm{k}}$ and $1/\it{g}^{\gamma}_{\bm{k},i\omega_m}=i\omega_m-\epsilon^{\gamma}_{\bm{k}}$.
}.
The electron-nematic phase transition shown here is described by a spontaneous distortion of the Fermi surface, caused by $C_3$-breaking hopping anisotropy. In addition, due to the symmetry of spin and valley, we perform a mean-field analysis dealing with all spin and valley degrees of freedom on an equal footing. In the following section, without loss of generality, we focus on one-spin and one-valley degrees of freedom.
We summarize the mean-field phase diagram ($T$, $N$), with the temperature $T$ and the filling $N$, determined by the Landau free energy in Fig. \ref{fig:phase}.
The transition is of purely second-order at VH filling ($N_{\rm VH}\sim1.25$) because of $u_3=0$ and $u_4>0$. We note that $N=2$ corresponds to the full filling and $N=1$ corresponds to the charge-neutral point.
The important feature is that the transition is of weak first-order in a wide range of filling.
``Weak first-order'' means that the character of the phase transition is first order but the transition temperature is close to the second-order transition temperature, which is defined by $u_2=0$. 
In general, first-order transitions are not accompanied by a divergence of the susceptibility, but a remnant of critical fluctuations can nevertheless be observed due to the vicinity of the second order instability, as we will show below.

We show the temperature dependence of the order parameters in Fig. \ref{fig:nem}. We note that a finite value of the order parameter yields a deformation of the Fermi surface which breaks the $C_{3z}$ symmetry.
Although, in the vicinity of VH filling, the transition is of almost second-order with a continuous change of the order parameter in Fig. \ref{fig:nem}(a), for other fillings, the transition is of weak first-order with a small discontinuous change of the order parameter in Fig. \ref{fig:nem}(b).
In this weak first-order region, we expect a nearly diverging behavior of the nematic susceptibility.
See Appendix \ref{appendix:hartreefock} for details about changes of DOS, band structure, and Fermi surface.

\begin{figure}
\includegraphics[width=7cm]{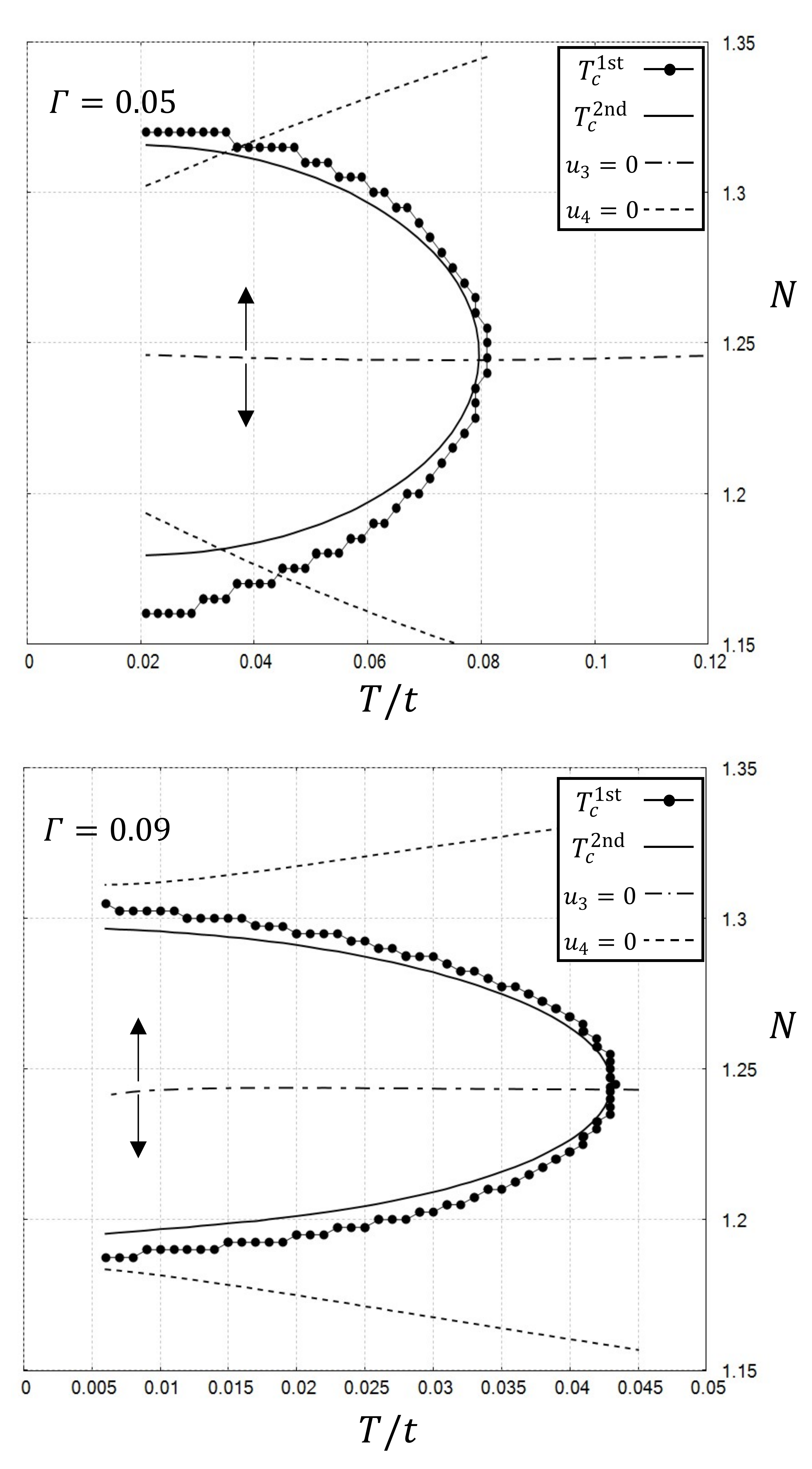}
\caption{
Phase diagrams of a bond-ordered phase with the impurity scattering ($\Gamma=0.05$ and $\Gamma=0.09$).
$T_c^{1{\rm st}}$ ($T_c^{2{\rm nd}}$) is the first (second)-order phase transition point, and $u_3=0$ ($u_4=0$) is zeros of $u_3$ ($u_4$).
The calculation is done by using a square mesh of $100\times 100$ in the Brillouin zone and a mesh of $1000$ in the energy. 
The phase transition line $T_c^{1{\rm st}}$ is defined by $F_{\rm{nem}}[\Phi_1, \Phi_2]=0$ and $\partial F_{\rm{nem}}[\Phi_1, \Phi_2]/\partial \Phi_i =0$ with $i=1,2$.
}
\label{fig:phase_imp}
\end{figure}

Next, we show how weak impurity scattering modifies the mean-field phase diagram. 
In graphene-based materials, it is known that there are impurity effects due to the substrate and disorder effects due to sample inhomogeneity. Here, for simplicity, we treat the impurity effect at the level of the Born approximation introduced in Eq. (\ref{eq:born_selfenergy}). 
In Fig. \ref{fig:phase_imp}, the mean-field phase diagrams for disordered cases ($\Gamma=0.05$ and $\Gamma=0.09$) are shown. First, we observe that the transition temperature of the three-state Potts nematic state is suppressed with increasing the impurity scattering. Second, the first order transition line at low temperatures gradually approaches the second-order one, rendering the transition a weak first-order. Thus we conclude that the transition becomes weakly first-order in the presence of the weak impurity scattering.

As described above, we have used the mean-field approximation for the free energy and the critical properties.
In general, it is known that phase transitions and critical properties can be modified by introducing mode-coupling effects between fluctuations, such as third- and fourth-order terms of GL action. In addition, due to the peculiarities of the three-state Potts model, the classical phase transition at finite temperature is known to be a second-order transition in two spatial dimensions\cite{Baxter_JOPC1973,Wu_RMP1982}, and it is expected that the first-order transition discussed here will be closer to the second-order transition if we take into account the mode-coupling effect\cite{Hertz_PRB1976,Millis_PRB1993,SCR,Wolfle_RMP2007}. Of course, in the case of quantum phase transitions\cite{Li_Nat2017,ZBi_PRR2020,YXu_PRB2020}, the order of the phase transition is not well understood, and it is an open question what happens for the order of the phase transition when the nematic phase transition is accompanied by loop-current order or when the impurity vertex corrections are applied.
In our analysis, the critical properties near the phase transition point are due to the mean-field approximation, but the stability and the extent of the ordered phase are expected to remain qualitatively unchanged even if the effects of such fluctuations are included.

Before closing this subsection, we comment on the connection between the calculation and experimental observations.
In Ref. \onlinecite{Cao_AX2020_NSC}, the authors obtained the phase diagram by changing the filling with a gate voltage, where the electron-nematic state is realized only in a narrow filling range. This observation is consistent with the fact that the electron-nematic state is stable only near the VHs in our mean-field calculations.

\subsection{Sound attenuation coefficients}
\begin{figure}
\includegraphics[width=8cm]{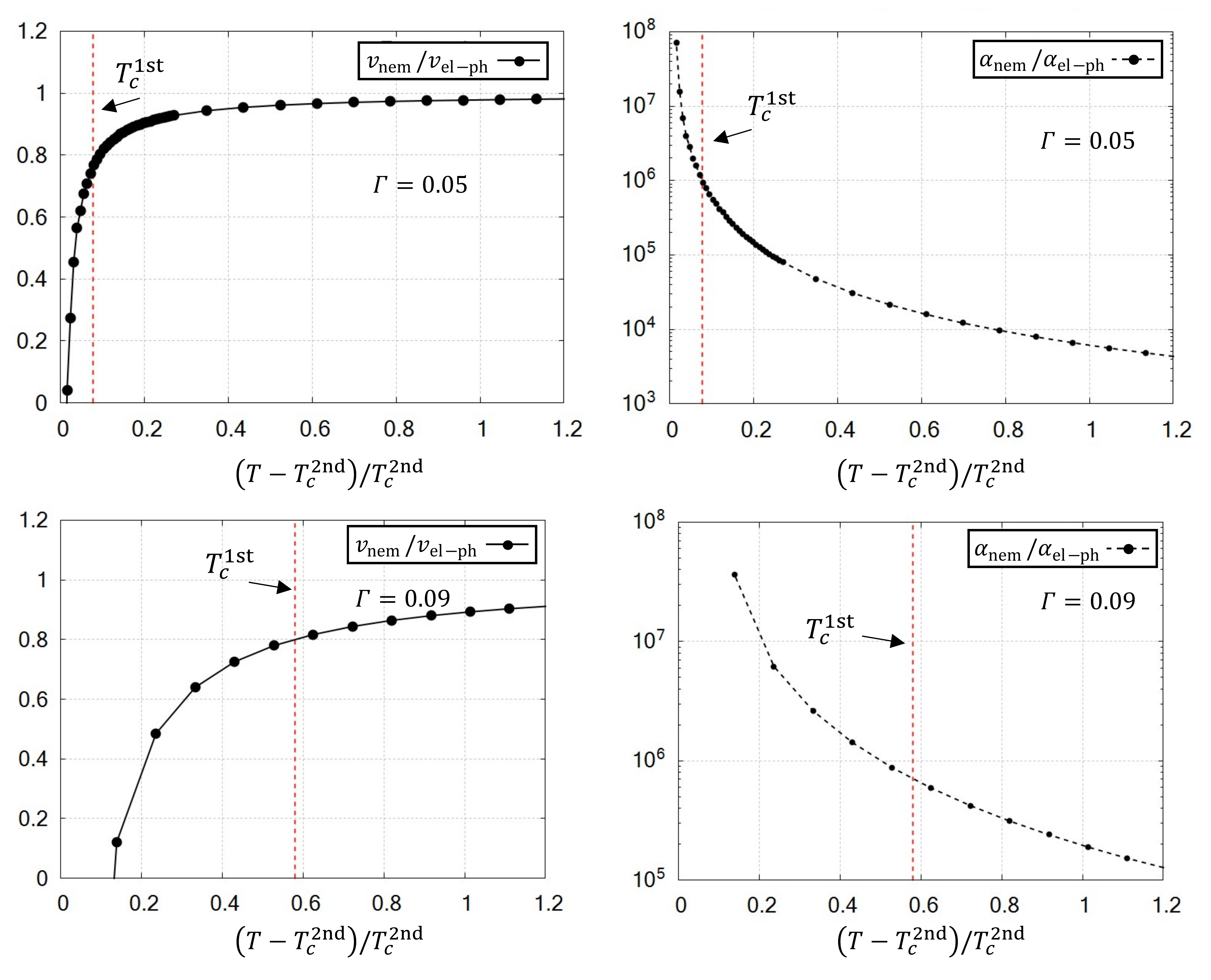}
\caption{
Temperature dependence of the sound velocities $v_{\rm nem}/v_{\rm ep-ph}$ and the sound attenuation coefficients $\alpha_{\rm nem}/\alpha_{\rm ep-ph}$ for the transverse acoustic wave for $N=1.2$. $T_c^{1{\rm st}}$ ($T_c^{2{\rm nd}}$) is the first (second) order phase transition point. The points only make sense above the transition temperature for $T>T_c^{1{\rm st}}$.
}
\label{fig:coeff}
\end{figure}

Next, we show the sound attenuation coefficients and the sound wave velocity for the transverse acoustic phonons, which are modified by the Fermi surface fluctuation.
The phonon self-energy [$\delta K_{\mu, \rm{el-ph}}(q)$ due to the electron-phonon couplings and $\delta K_{\mu, \rm{nem}}(q)$ due to the nemato-elastic couplings in Eq. (\ref{eq:phonon_full})] are obtained numerically. 
Using these self-energies, we calculate the normalized sound velocities $v_{\rm nem}/v_{\rm el-ph}$ and the normalized sound attenuation coefficients $\alpha_{\rm nem}/\alpha_{\rm el-ph}$, which quantify the contribution of the nematic fluctuation ($v_{\rm nem}$, $\alpha_{\rm nem}$) to the electron-phonon coupling ($v_{\rm el-ph}$, $\alpha_{\rm el-ph}$).
The temperature dependencies of the transverse sound velocity and the transverse sound attenuation coefficient for several impurity scatterings are shown in Fig. \ref{fig:coeff}. 
The parameter region is in the weak first-order phase transition for $N=1.2$.
We note that the ratio of the sound velocity $v_{\rm nem}/v_{\rm el-ph}$ takes about 0.8 at $T_c^{1{\rm st}}$ for the choice of parameters.

It is confirmed that the ultrasound attenuation coefficient is enhanced by a factor of about 100 around the first-order transition temperature $T_c^{1 st}$ even if the impurity effect is present in Fig. \ref{fig:coeff} ($\Gamma=0.05$).
Furthermore, in the region where the impurity scattering is much stronger in Fig. \ref{fig:coeff} ($\Gamma=0.09$), the ultrasound attenuation coefficient is still enhanced by a factor of 10 for the same parameters as above. These results suggest that the weak first-order phase transition occurs and that the effect of nematic fluctuations can be observed in the phonon damping even in the presence of impurities.

\section{Discussion}
Here some additional comments are in order on the characteristic properties discussed in the previous sections.

{\it{Superlattice effects:}}
In this paper, we focus on the long-wavelength limit of acoustic phonons with linear dispersions. Here we comment on the phonon modes in MA-TBG, which are complicated due to the superlattice structure. 
One of the unique properties of such moir\'e phonon modes, which reflects a non-rigid crystal\cite{Koshino_PRB2019,Ochoa_PRB2019}, is the appearance of rotation tensors in addition to the ordinary strain tensors in the elastic degrees of freedom.
While for a rigid crystal the velocity of longitudinal phonons is much larger than that of transverse phonons, for twisted bilayer graphene having a non-rigid crystal property, the velocity of transverse phonons may exceed that of longitudinal phonons due to the coupling between the strain tensor and the rotation tensor\cite{Koshino_PRB2019,Ochoa_PRB2019,Fernandes_AdvSci2020}.  
Although there are such quantitative differences, as far as the acoustic phonons in the long-wavelength limit are concerned, there is no qualitative change in their linear-dispersion properties. 
Thus we believe, even in the moir\'e materials, that our theory can be applied to the low-energy properties of acoustic phonons with linear dispersions.
It is also important to consider the effect of the rotation tensor to the nematicity as discussed in Ref. \onlinecite{Fernandes_AdvSci2020}. Since the electron-nematic order parameter does not couple to the rotation tensor in the leading order correction, we have not considered it in this paper. Nevertheless, the detailed study including the higher order corrections to phonons and nematicity is important; this is left for future work.

{\it{Impurity effects:}} 
In addition to the discussions in Secs. II. C and III. C, it is also important to consider several scattering mechanisms. In this connection, we comment here on the impurity effects beyond the Born approximation in Eq. (\ref{eq:born_selfenergy}). 
We expect that the impurity vertex correction changes the dynamical critical exponent, leading to the change of the wavenumber dependence of the ultrasound attenuation coefficients from $|\bm{q}|$ to $|\bm{q}|^2$.

The damping rate $\Gamma_d(q)$ in Eq. (\ref{eq:gaussian_fluctuation}) depends on the dynamical critical exponent $z$ as,
\begin{eqnarray}
\Gamma_d(\bm{q})&=&\gamma_d |\bm{q}|^{z-2},
\end{eqnarray}
where $z=3$ for a usual ferroic order in clean systems.
One of the unique properties of the electron-nematic state is that the sound attenuation coefficient in Eq. (\ref{eq:sound_attenuation}) reflects the damping rate of nematic fluctuations, as follows;
\begin{eqnarray}
\alpha_{\mu}(\bm{q})&\sim& \frac{\kappa^2 }{2\rho v^*_{\mu}}
\frac{1}{r^2 }\frac{\bm{q}^2}{\Gamma_d(\bm{q})}= \frac{\kappa^2 }{2\rho v^*_{\mu}}
\frac{1}{r^2 }\frac{|\bm{q}|^{4-z}}{\gamma_d}.
\end{eqnarray}

We discuss how the impurity effect would modify the above nematic fluctuations via a possible change in the exponent $z$.
It is known that for charge density fluctuations, a diffusion pole appears from vertex corrections for the impurity scattering\cite{PALee_RMP1985,Belitz_RMP1994,Belitz_RMP2005,Wolfle_RMP2007}
, and the dynamical critical exponent becomes $z=4$. This is related to the conservation law of electric charge, and such a diffusive mode appears when there is charge $U(1)$-gauge symmetry.
On the other hand, in the present case of electric quadrupoles (the electric quadrupole density is not a conserved quantity\cite{Gallais_CRPhys2016,Mattia_PRL2020}), it is expected that the normal diffusion mode does not appear due to impurity effects\cite{Klein_PRR2019}, and we expect the relaxation mode\cite{Millis_PRL2002} with $z=2$, etc. 
In this case, the dynamical critical exponent may be changed to a value other than $z=3$, unlike the usual charge density fluctuation, and this change will be probed through the wave-number dependence of the ultrasound attenuation coefficient. 
To identify the correct dynamical exponent is an open problem, and further analysis will be required.

{\it{Candidate materials for experiments:}}
A three-state Potts nematic order has been reported for doped-Bi$_2$Se$_3$\cite{Kuntsevich_NJP2018,YSun_PRL2019,Cho_Nat2020}.
Even in these materials, as the 2D nematic ordered state which breaks the in-plane $C_{3z}$-symmetry is stacked in the $z$-direction, the formulation developed here can be applied to phonon modes propagating in the plane with a slight modification.
In these materials, it has been suggested that a vestigial nematic order\cite{Hecker_npj2018} is caused by nematic superconducting fluctuations, rather than the bond-order discussed here.
Nevertheless, a similar treatment can be applied, and thus we expect the isotropic divergence of sound attenuation and the isotropic lattice softening for transverse modes within the GL theory discussed here. The scenario presented here is useful to probe the nematic fluctuation, predicting a weak first-order transition like behavior.

In the case of MA-TBG, an electron-nematic state has been reported at several fillings by scanning tunneling microscopy\cite{YChoi_NatPhys2019,AKerelsky_Nat2019,YJiang_Nat2019}, transport measurement\cite{Cao_AX2020_NSC}, and the quantum oscillation\cite{Cao_AX2020_NSC,SLiu_AX2019}. 
Our mean-field analysis for the $C_3$-breaking bond order is based on Ref. \onlinecite{Onari_AX2020}. 
It is shown that the $C_3$-breaking intra-valley bond ordered state is stabilized near the VH filling, and the other magnetically ordered states are suppressed by using the so-called DW equation method including the Aslamazov-Larkin vertex correction\cite{Onari_AX2020}.
Besides the weak-coupling approaches\cite{Onari_AX2020,Chichinadze_PRB2020}, there are some theoretical proposals such as an orbital order and a vestigial nematic order in the strong coupling theory\cite{Venderbos_PRB2018}.
We think that our phenomenological theory can also be applied to the above scenarios with a slight modification. Detailed study on this point is left for future work.

Unfortunately, MA-TBG does not allow us to conduct usual sound attenuation experiments due to its purely 2D character, but this does not change the fact that the mean-free path $l=\alpha^{-1}$ of phonons becomes isotropically shorter. 
In this 2D case, experiments using optical methods such as Brillouin scattering\cite{ZKWang_Carbon2008} and double resonant Raman scattering\cite{XCong_Carbon2019} provide alternative probes to detect the nematic fluctuation. For these experiments, the formulation developed here can be applied with a slight modification to identify such a three-state Potts nematic state and figure out whether it is induced spontaneously or from trivial strain.

\section{SUMMARY}

We have analyzed the impact of nemato-elastic coupling on the low-energy properties of phonons by using a phenomenological argument and a model calculation. 
Phenomenological analysis has clarified that the Landau damping term becomes isotropic due to fluctuations of the $C_3$-breaking bond-order in the Gaussian fluctuation region, and the nemato-elastic coupling is also isotropic. 
As a result, we have proposed to detect the intrinsic three-state Potts nematic phase transition by measuring the ultrasound attenuation of the transverse acoustic phonon.
Namely, the ultrasound attenuation coefficient shows an isotropic divergence which is proportional to the momentum $|\bm{q}|$, and the sound velocity renormalization also shows an isotropic angle dependence. 
Both features are quite contrasted to the strong anisotropy in the case of the $C_4$-breaking nematic case.

We have determined the phase diagram by using an extended Hubbard model in a mean-field approximation to investigate the critical properties.
According to the mean-field approximation, the transition temperature takes its maximum near VHs, and the large density of states favors the nematic phase transition. The order of phase transition is of weak first-order in a wide range of band filling and, with increasing the impurity scattering, the first order transition line at low temperatures gets closer to the second-order line, making the transition weakly first-order in a wider parameter region.
Furthermore, it has been confirmed that the enhancement of the ultrasound attenuation coefficient can be observed in the case of a weak first-order phase transition. 
Even if the effect of mode coupling between the nematic fluctuations is considered, the qualitative features of the isotropic sound attenuation coefficients and the phase diagram are expected to be unchanged, but the order of the transition could be changed to the second-order as expected for a classical phase transition of three-state Potts nematicity in 2D.

\begin{acknowledgments}
We are grateful to S. Sumita for his helpful contribution in the early stage of this work.
We would like to thank R. Toshio, H. Watanabe, K. Adachi, K. Takasan, H. Adachi and Y. Yanase for fruitful discussions and useful comments.
This work was partly supported by JSPS KAKENHI (Grant No. 20J13688, JP19H01838, JP18H01140).
K. K. is supported by WISE Program from MEXT, and a Research Fellowship for Young Scientists from JSPS.
M.S. is grateful for the support by the Swiss National Science 
Foundation (SNSF) through Division II (No. 184739).
The numerical calculations were performed on the supercomputer at the Institute for Solid State Physics in the University of Tokyo.
\end{acknowledgments}

\appendix

\makeatletter
	\renewcommand{\thefigure}{
	\thesection.\arabic{figure}}
	\@addtoreset{figure}{section}

	\renewcommand{\thetable}{
	\thesection.\arabic{table}}
	\@addtoreset{table}{section}
\makeatother

\section{Nematic Polarization for a circular Fermi surface \label{appendix:nematicpolarization}}

Here, we derive the functional form of the Landau damping $\hat{D}\Bigl(\frac{|\epsilon_m|}{\Gamma_{d}(\bm{q})}\Bigr)$ in Eq. (\ref{eq:nemati_propargator}), which results in Eq. (\ref{eq:gaussian_fluctuation}).
We assume the circular Fermi surface around the $\Gamma$ point and the single band system in a $C_3$ symmetric lattice.
The interaction between the nematic fluctuation ($\Phi_{1\bm{q}},\Phi_{2\bm{q}},$) and the electrons ($c^{\dagger}_{\bm{k}},c_{\bm{k}}$) is given by 
\begin{eqnarray}
\mathcal{H}_{\rm coup}&\propto &\sum_{\bm{q},\bm{k}}
\Bigl[d_{1\bm{k}}\Phi_{1\bm{q}}+d_{2\bm{k}}\Phi_{2\bm{q}}\Bigr]
c^{\dagger}_{\bm{k}+\bm{q}/2}c_{\bm{k}-\bm{q}/2},
\end{eqnarray}
where form factors of a two-dimensional representation are $d_{1\bm{k}} \sim (\hat{k}_x^2-\hat{k}_y^2)=\cos{2\theta_{\bm{k}}}$ and $d_{2\bm{k}}\sim 2(\hat{k}_x \hat{k}_y)=\sin{2\theta_{\bm{k}}}$ with the wave vector of electron $\bm{k}=|\bm{k}|(\hat{k}_x,\hat{k}_y)=|\bm{k}|(\cos{\theta_{\bm{k}}},\sin{\theta_{\bm{k}}})$. 
Furthermore, the order parameter is parametrized as $\bm{\Phi}=\Phi(\cos{2\theta},\sin{2\theta})$ with the nematic director $\hat{n}=(\cos{\theta},\sin{\theta})$ and its angle $\theta$. Thus, the coupling term is expressed in terms of the relative angle $\theta_{\bm{k}}-\theta$ as follows:
\begin{eqnarray}
\mathcal{H}_{\rm coup}&\propto &\sum_{\bm{q},\bm{k}}
\Phi_{\bm{q}}\cos{2(\theta_{\bm{k}}-\theta)}
c^{\dagger}_{\bm{k}+\bm{q}/2}c_{\bm{k}-\bm{q}/2}, 
\end{eqnarray}
where the coupling term vanishes at $\theta_{\bm{k}}-\theta=\pm \pi/4$.  

The low-energy contribution of a nematic polarization $\chi_{q}^{ij}$ determines the dynamical properties of the nematic fluctuations. The $\bm{k}$-summation can be performed by linearizing the electronic dispersion, 
\begin{eqnarray} 
\chi_{q}^{ij}&=&\sum_{k} d_{i\bm{k}}d_{j\bm{k}} G_{k}G_{k+q} \sim -i\epsilon_m
\rho_{0} \int_{\bm{k}_{\rm FS}} \frac{d_{i\bm{k}}d_{j\bm{k}} }{i\epsilon_m-v_{\rm F}\bm{k}\cdot \bm{q}},\nonumber \\
&=&-\frac{i\epsilon_m}{v_{\rm F}|\bm{q}|}\rho_{0}
\int_0^{2\pi}\frac{d\theta_{\bm{k}}}{2\pi}
\frac{d_{i\bm{k}}d_{j\bm{k}} }{i\epsilon_m/v_{\rm F}|\bm{q}|-\cos{(\theta_{\bm{k}}-\theta_{\bm{q}})}}, \nonumber \\
\end{eqnarray}
with $d_{1\bm{k}}d_{1\bm{k}}=\cos^2{2 \theta_{\bm{k}} }$, 
$d_{2\bm{k}}d_{2\bm{k}}=\sin^2{2\theta_{\bm{k}}}$, $d_{1\bm{k}}d_{2\bm{k}}=\sin{2\theta_{\bm{k}}}\cos{2\theta_{\bm{k}}}$, $\rho_{0}$ is the density of states at the Fermi level, an electron Green's function $G_k^{-1}=i\omega_n-\epsilon_{\bm{k}}-\mu$, the energy dispersion $\epsilon_{\bm{k}}=\bm{k}^2/2m$, the electron mass $m$, the Fermi velocity $v_{\rm F}$, the fermion Matsubara frequency $\omega_n$, and the boson Matsubara frequency $\epsilon_m$.
Now we set $\psi=(\theta_{\bm{k}}-\theta_{\bm{q}})$ and rewrite each component of $d_{i\bm{k}}d_{j\bm{k}}$ as,  
\begin{eqnarray}
d_{1\bm{k}}d_{1\bm{k}}&=&\cos^2{2 \theta_{\bm{k}} }= \cos^2{(2 \psi+2 \theta_{\bm{q}})}\nonumber \\
&\sim& \cos^2{2\theta_{\bm{q}}}\cos^2{2 \psi} +\sin^2{2\theta_{\bm{q}}}\sin^2{2 \psi}  ,\\
d_{2\bm{k}}d_{2\bm{k}}&=&\sin^2{2\theta_{\bm{k}}}= \sin^2{(2 \psi+2 \theta_{\bm{q}})}\nonumber \\
&\sim&\cos^2{2\theta_{\bm{q}}} \sin^2{2 \psi}+\sin^2{2\theta_{\bm{q}}} \cos^2{2 \psi}  ,\\
d_{1\bm{k}}d_{2\bm{k}}&=&\frac{1}{2}\sin{4 \theta_{\bm{k}}}= \frac{1}{2} \sin{(4 \psi+4 \theta_{\bm{q}})}
\nonumber \\
&\sim&\frac{1}{2} \cos{4 \theta_{\bm{q}}}\sin{4 \psi}+\frac{1}{2} \sin{4 \theta_{\bm{q}}}\cos{4 \psi},\nonumber \\
&=&\frac{1}{2} \sin{4 \theta_{\bm{q}}}\bigl( 2\cos^2{2 \psi}-1\bigr) ,
\end{eqnarray}
where we have ignored terms proportional to $\sin{2 \psi}\cos{2 \psi}$, because they vanish after $\psi$ integral. Combined with the above equations, the dynamical part of nematic polarization $\hat{D}_q=\hat{\chi}_q-\hat{\chi}_{\bm{q},0}$ is calculated as, 
\begin{eqnarray}
D_{q}^{ij}&=&-ia \rho_{0}
\int_0^{2\pi}\frac{d\psi}{2\pi}
\frac{d_{i\bm{k}}d_{j\bm{k}} }{ia-\cos{\psi}}, 
\end{eqnarray}
with $a=\frac{\epsilon_m}{v_{\rm F}|\bm{q}|}$ and 
\begin{eqnarray}
D_{q}^{11}&=&ia \rho_{0}
\Bigl\{
\cos^2{2\theta_{\bm{q}}}
\Bigl[i{\rm Sgn}(a)-2i a\Bigr]
+\sin^2{2\theta_{\bm{q}}}
\Bigl[2i a\Bigr]
\Bigr\}, \nonumber \\ \\
D_{q}^{22}&=&ia \rho_{0}
\Bigl\{
\cos^2{2\theta_{\bm{q}}}
\Bigl[2i a\Bigr]
+\sin^2{2\theta_{\bm{q}}}
\Bigl[i{\rm Sgn}(a)-2i a\Bigr]
\Bigr\},\nonumber \\ \\
D_{q}^{12}&=&ia \rho_{0}
\Bigl\{
\sin{4\theta_{\bm{q}}}
\Bigl[i{\rm Sgn}(a)-2i a\Bigr]
-\frac{1}{2}\sin{4\theta_{\bm{q}}}
\Bigl[i{\rm Sgn}(a) \Bigr]
\Bigr\}, \nonumber \\
\end{eqnarray}
with $D_{q}^{12}=D_{q}^{21}$.
We have used the following equations in the above calculations:
\begin{eqnarray}
I_C(a)&=&-\int_0^{2\pi}\frac{d\psi}{2\pi}
\frac{\cos^2{2\psi}}{ia-\cos{\psi}}, \nonumber \\
&=&i(1+2a^2)\Bigl[ \frac{(1+2a^2)}{\sqrt{1+a^2}}{\rm Sgn}(a)-2a \Bigr]|_{a\rightarrow 0},
\nonumber \\ &&\rightarrow i{\rm Sgn}(a)-2i a,\\
I_S(a)&=&-\int_0^{2\pi}\frac{d\psi}{2\pi}
\frac{\sin^2{2\psi}}{ia-\cos{\psi}}, \nonumber \\
&=&2ai\Bigl[1+2a^2-2|a|\sqrt{1+a^2} \Bigr]|_{a\rightarrow 0},
\nonumber \\ &&\rightarrow 2i a.
\end{eqnarray}
As a consequence, the frequency-dependent part of nematic polarization is given as,
\begin{eqnarray}
\hat{D}_{q}
&=&
-\frac{a\rho_{0}}{2}
\left(
\begin{array}{cc}
1& 0 \\
0 & 1
\end{array}
\right)\nonumber \\
&&-\rho_{0}\Bigl[ \frac{a}{2}-2a^2\Bigr]
\left(
\begin{array}{cc}
\cos{4\theta_{\bm{q}}}& \sin{4\theta_{\bm{q}}}\\
\sin{4\theta_{\bm{q}}} & -\cos{4\theta_{\bm{q}}}
\end{array}
\right).\nonumber \\
\end{eqnarray}
Next, we express the above function in terms of the angle of nematic director $\theta$ and consider the dynamical part of Eq. (\ref{eq:gauss}), 
\begin{eqnarray}
\bm{\Phi}^T_{q}\hat{D}_{q}\bm{\Phi}^*_{q}
&=&
\Phi_q
\left(
\begin{array}{cc}
\cos{2\theta} & \sin{2\theta} 
\end{array}
\right)
\hat{D}_{q}
\left(
\begin{array}{c}
\cos{2\theta} \\  \sin{2\theta} 
\end{array}
\right)\Phi^*_q,\nonumber \\
&=&
-\Phi_q
\rho_{0}\Bigl[
\frac{|\epsilon_m|}{v_{\rm F}|\bm{q}|}\cos^2{(2\theta_{\bm{q}}-2\theta)}\nonumber \\
&&-2\frac{|\epsilon_m|^2}{(v_{\rm F}|\bm{q}|)^2} \cos{(4\theta_{\bm{q}}-4\theta)}\Bigr]\Phi^*_q,
\end{eqnarray}
where $\bm{\Phi}_q=\Phi_q(\cos{2\theta},\sin{2\theta})$ and $\Phi_q$ is the norm of $\bm{\Phi}_q$.
Thus the Gaussian theory for the three-state Potts nematic fluctuation is described by  
$S_{\rm{Gauss}}[\Phi]=\int_q \Phi_q \Bigl[ r+\xi_0^2\bm{q}^2-D_q \Bigr]\Phi^*_q$ with 
\begin{eqnarray}
D_q&=&-
\rho_{0}\Bigl[
\frac{|\epsilon_m|}{v_{\rm F}|\bm{q}|}\cos^2{(2\theta_{\bm{q}}-2\theta)}\nonumber \\
&&-2\frac{|\epsilon_m|^2}{(v_{\rm F}|\bm{q}|)^2} \cos{(4\theta_{\bm{q}}-4\theta)}\Bigr],
\end{eqnarray}
where $r\propto T_{c0}-T$ measures the distance from the mean-field transition temperature $T_{c0}$, where the mean-field correlation length is $\xi_0$.
The orientations of the nematic directors are restricted to three directions by the cubic term in Eq. (\ref{eq:GL_action}) as follows: 
$\theta=\{0, 2\pi/3, 4\pi/3\} $ for $u_3<0$ and $\theta=\{-\pi/6, \pi/2, 7\pi/6\}$ for $u_3>0$. 
Precisely speaking, the damping term preserves this $\mathbb{Z}_3$ symmetry in a disordered state, thus we need to treat three angles equivalently,
\begin{eqnarray}
&&\cos^2{(2\theta_{\bm{q}}-2\theta)}\nonumber \\
&\rightarrow& \frac{1}{3}\Bigl[ \cos^2{(2\theta_{\bm{q}})}+\cos^2{(2\theta_{\bm{q}}-\frac{2\pi}{3})}\nonumber \\
&&+\cos^2{(2\theta_{\bm{q}}-\frac{4\pi}{3})}\Bigr],\\
&=&\frac{1}{3}\Bigl[ \cos^2{(2\theta_{\bm{q}})}+ \frac{1}{2}\cos^2{(2\theta_{\bm{q}})}+\frac{3}{2}\sin^2{(2\theta_{\bm{q}})}\Bigr],\nonumber \\&=&\frac{1}{2}.
\end{eqnarray}
Therefore, within this treatment, there is no anisotropy of Landau damping in the three-state Potts nematic case, and thus we can use the following action,
\begin{eqnarray}
S_{\rm{Gauss}}[\Phi]&=&\int_q \Phi_q \Bigl[ r+\xi_0^2\bm{q}^2-D_q \Bigr]\Phi_q^*, \nonumber \\ 
D_q&=&-
\frac{\rho_{0}}{2}
\frac{|\omega_m|}{v_{\rm F}|\bm{q}|},
\end{eqnarray}
as shown in Eq. (\ref{eq:gaussian_fluctuation}).

\section{Nemato-elastic coupling \label{appendix:nematoelasticcoupling}}

Here, we derive the nemato-elastic coupling in Eq. (\ref{eq:nemato_elastic_coupling}).
In terms of $\tilde{u}_{L}(q)$ and  $\tilde{u}_{T}(q)$, the nemato-elastic action reads
\begin{eqnarray}
S_{\rm{nem-ph}}[\bm{\Phi},\bm{u}]&=&-\kappa \int_q 
\left(
\begin{array}{cc}
\tilde{u}_{L}(q) & \tilde{u}_{T}(q)
\end{array}
\right)
\nonumber \\
&\times&
i|\bm{q}|
\left(
\begin{array}{cc}
\cos{2\theta_{\bm{q}}}& \sin{2\theta_{\bm{q}}} \\
-\sin{2\theta_{\bm{q}}} & \cos{2\theta_{\bm{q}}}
\end{array}
\right)
\left(
\begin{array}{c}
\Phi_1(-q) \\
\Phi_2(-q)
\end{array}
\right), \nonumber \\
&=&-\kappa  \int_q
\left(
\begin{array}{cc}
\tilde{u}_{L}(q) & \tilde{u}_{T}(q)
\end{array}
\right)
\nonumber \\
&\times&  i|\bm{q}|
\left(
\begin{array}{c}
\cos{(2\theta_{\bm{q}}-2\theta)} \\ 
-\sin{(2\theta_{\bm{q}}-2\theta)} 
\end{array}
\right)
\Phi^*(q), \nonumber \\
\end{eqnarray}
where $\theta_{\bm{q}}$ shows the propagating direction of a wave vector $\bm{q}$.
In the second line, we have used $\bm{\Phi}=\Phi(\cos{2\theta},\sin{2\theta})$.

In the case of the Ising nematicity, the nematic director is forced to be $\theta=\{0,\pi/2 \}$ for $d_{x^2-y^2}$-wave. 
Even if we treat the two angles equally, the anisotropy of the nemato-elastic coupling remains, as follows,
\begin{eqnarray}
\cos^2{(2\theta_{\bm{q}}-2\theta)}&\rightarrow& \cos^2{(2\theta_{\bm{q}})},\\ 
\sin^2{(2\theta_{\bm{q}}-2\theta)} &\rightarrow& \sin^2{(2\theta_{\bm{q}})}.
\end{eqnarray}
This form is the same as in Ref. \onlinecite{Paul_PRL2017}.
However, in the case of the three-state Potts nematicity, treating the three angles equally does not show any anisotropy. Thus we conclude that $\mathbb{Z}_3$ symmetry leads to an isotropic angular dependence of the nemato-elastic coupling.
\begin{eqnarray}
\cos^2{(2\theta_{\bm{q}}-2\theta)}&\rightarrow& \frac{1}{2},\\ 
\sin^2{(2\theta_{\bm{q}}-2\theta)} &\rightarrow& \frac{1}{2}.
\end{eqnarray}
Thus we obtain the following isotropic form:
\begin{eqnarray}
S_{\rm{nem-ph}}[\Phi,\bm{u}]&=&-\kappa  \int_q  i\frac{|\bm{q}|}{2}
\Bigl[ \tilde{u}_{L}(q) -\tilde{u}_{T}(q)\Bigr]\Phi^*(q). \nonumber \\
\end{eqnarray}

\section{Quadrupole-Quadrupole interaction\label{appendix:interaction}}

\begin{figure}
\includegraphics[width=9cm]{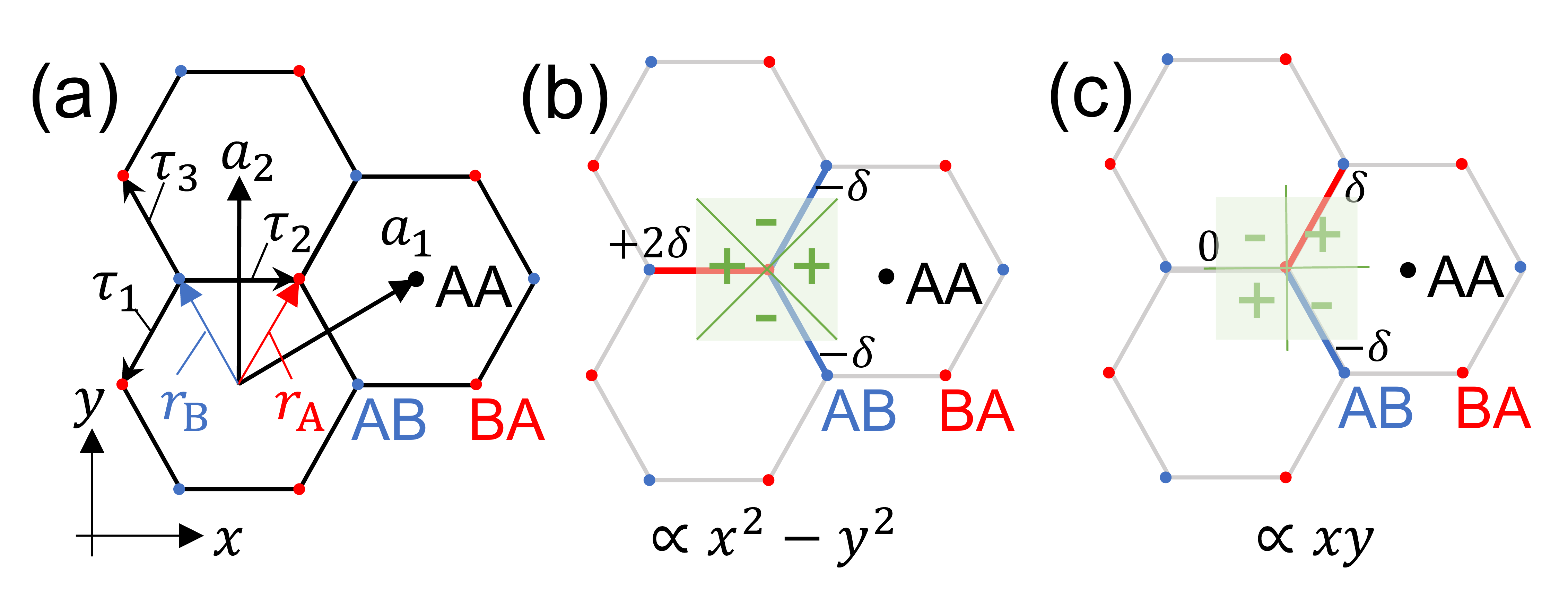}
\caption{
(a) The primitive lattice vectors on the honeycomb lattice: $\bm{a}_1=(\frac{\sqrt{3}}{2},\frac{1}
{2})$, $\bm{a}_2=(0,1)$ with vectors of the nearest-neighbor bond $\bm{\tau}_1=(-\frac{1}{2\sqrt{3}},-\frac{1}{2})$, $\bm{\tau}_2=(\frac{1}{\sqrt{3}},0)$, and $\bm{\tau}_3=(-\frac{1}{2\sqrt{3}},\frac{1}{2})$.
 The basis function of $E$ (or $E_g$) representation of form factor: (b) d$_{x^2-y^2}$-wave, (c) d$_{xy}$-wave}.
\label{fig:form}
\end{figure}

Here, we derive the forward-scattering interaction\cite{Yamase_PRB2007,HYKee_PRB2003,Khavkine_PRB2004,Yamase_PRB2005,Valenzuela_NJP2008} in Eq. (\ref{eq:forward_scattering}).
We note the atomic structure of TBG. In a small twist angle TBG, a slight mismatch in the lattice periods of two graphene layers gives rise to a long-period moir\'e interference pattern.
The regions that locally appear to be $AB$-stacked bilayer grahene and $BA$-stacked bilayer graphene form the emergent honeycomb lattice in Fig.\ref{fig:form}(a). 
Furthermore, it is pointed out that the Wannier state\cite{JKang_PRX2018_Wannier,MKoshino_PRX2018_Wannier,Zou_PRB2018_Wan} is centered at the $AB$ or $BA$ spot in the moir\'e pattern, while the maximum amplitude is at three $AA$ spots. 
Because of the three-peak form of the Wannier state, the Coulomb interaction between the neighboring sites is as important as the on-site interaction\cite{MKoshino_PRX2018_Wannier}.
Considering the nearest-neighbor (NN) direct channel on the multi-orbital Hubbard model, the interaction term is given by 
\begin{eqnarray}
H_{\rm{int}}&=&
\frac{1}{2}\sum_{a b}\sum_{\sigma \sigma^{\prime}}V_{ab}
c^{\dagger}_{a\sigma}
c_{a\sigma}
c_{b\sigma^{\prime}}^{\dagger}
c_{b\sigma^{\prime}},\\
&=&\frac{V_{\rm NN}}{2N}\sum_{\alpha \neq \beta}\sum_{\bm{q}}
\gamma_{\alpha \beta}^{\rm NN}(\bm{q})\rho_{\alpha}(\bm{q})\rho_{\beta}(-\bm{q}),\\
\gamma_{\rm{AB}, \rm{BA} }^{\rm NN}(\bm{q})&=&
\Bigl(e^{-i\bm{q}\cdot \bm{\tau}_1}+e^{-i\bm{q}\cdot \bm{\tau}_2}+e^{-i\bm{q}\cdot \bm{\tau}_3}\Bigr),
\end{eqnarray}
where $a=(i,\alpha,\xi)$ denotes the unit cell index $i$, the sublattice index $\alpha \in \{ \rm{AB}, \rm{BA}\}$, the valley index $\xi \in \{+,- \}$, and the density operator $\rho_{\alpha}(\bm{q})=\sum_{\alpha,\xi,\sigma}\sum_{\bm{k}}
c_{\bm{k}+\bm{q}\sigma}^{\alpha\xi\dagger}c_{\bm{k}\sigma}^{\alpha\xi}$.

We change the ordering of fermion operators in the NN direct channel as 
\begin{eqnarray}
&&\sum_{\bm{p}_1\bm{p}_2\bm{p}_3}\gamma_{\rm{AB}, \rm{BA} }^{\rm NN}(\bm{p}_2)
c_{\bm{p}_1+\bm{p}_2\sigma}^{\alpha \xi\dagger}c_{\bm{p}_1 \sigma}^{\alpha\xi}
c_{\bm{p}_3-\bm{p}_2\sigma^{\prime}}^{\beta \xi^{\prime}\dagger}c_{\bm{p}_3\sigma^{\prime}}^{\beta \xi^{\prime}} 
\nonumber \\&\sim&-\sum_{\bm{k}\bm{k}^{\prime}\bm{q}}
\gamma_{\rm{AB}, \rm{BA} }^{\rm NN}(\bm{k}-\bm{k}^{\prime})c^{ \alpha \xi\dagger}_{\bm{k}+\bm{q}/2 \sigma}c^{ \beta \xi}_{\bm{k}-\bm{q}/2 \sigma^{\prime}}
c^{ \beta \xi\dagger}_{\bm{k}^{\prime}-\bm{q}/2 \sigma}c^{\alpha \xi}_{\bm{k}^{\prime}+\bm{q}/2 \sigma^{\prime}},\nonumber \\
\label{eq:nninteraction}
\end{eqnarray}
where we have ignored the inter-valley component and consider only $\sigma =\sigma^{\prime}$.
Now we decouple $\gamma_{\rm{AB}, \rm{BA} }^{\rm NN}(\bm{k}-\bm{k}^{\prime})$ as $\gamma_{\rm{AB}, \rm{BA} }^{\rm NN}(\bm{k}-\bm{k}^{\prime})=\frac{1}{3}s^{*}_{\bm{k}}s_{\bm{k}^{\prime}}
+\frac{2}{3}E^{1*}_{\bm{k}}E^1_{\bm{k}^{\prime}}+\frac{2}{3}E^{2*}_{\bm{k}}E^2_{\bm{k}^{\prime}}$, where $s_{\bm{k}}$, $E^1_{\bm{k}}$, and $E^2_{\bm{k}}$ are the form factors in Figs.\ref{fig:form}(b) and \ref{fig:form}(c), such as 
\begin{eqnarray}
s_{\bm{k}}&=&e^{i\bm{k}\cdot \bm{\tau}_1}+e^{i\bm{k}\cdot \bm{\tau}_2}+e^{i\bm{k}\cdot \bm{\tau}_3},\nonumber  \\
E^1_{\bm{k}}&=&e^{i\bm{k}\cdot \bm{\tau}_1}-\frac{1}{2}e^{i\bm{k}\cdot \bm{\tau}_2}-\frac{1}{2}e^{i\bm{k}\cdot \bm{\tau}_3}, \nonumber \\
E^2_{\bm{k}}&=&-\frac{\sqrt{3}}{2}e^{i\bm{k}\cdot \bm{\tau}_2}+\frac{\sqrt{3}}{2}e^{i\bm{k}\cdot \bm{\tau}_3}.
\label{eq:form}
\end{eqnarray}
We can rewrite Eq. (\ref{eq:nninteraction}) in terms of the density operator, which is in the $E$-representation of the point group $D_3$, $n^{AB\xi \sigma}_{E_i}(\bm{q})=\frac{1}{N}\sum_{\bm{k}}E^{i*}_{\bm{k}}c^{AB\xi \dagger}_{\bm{k}+\bm{q}/2\sigma}c_{\bm{k}-\bm{q}/2\sigma}^{BA\xi}$, 
\begin{eqnarray}
H_{\rm{int}}^{AB}&=&-\frac{V_{\rm NN}}{3N}\sum_{\xi \sigma}\sum_{\bm{k},\bm{k}^{\prime},\bm{q}}
\Bigl(n^{AB \xi\sigma}_{E_1}(\bm{q})[n^{AB\xi\sigma}_{E_1}(\bm{q})]^{\dagger}\nonumber \\ 
&&+n^{AB\xi\sigma}_{E_2}(\bm{q})[n^{AB\xi\sigma}_{E_2}(\bm{q})]^{\dagger}\Bigr).
\label{eq:nninteraction_2}
\end{eqnarray}
Finally, we have $H_{\rm{int}}=H_{\rm{int}}^{AB}+H_{\rm{int}}^{BA}$ as shown in Eq. (\ref{eq:forward_scattering}).

In the $D_3$ point group case\cite{MKoshino_PRX2018_Wannier}, form factors result from TABLE.\ref{table:1}.
The real part of $n^{AB\xi \sigma}_{E_i}(\bm{q})$ corresponds to the $d$-wave components of the density operator, referred to as nematic fields, whereas the imaginary part of $n^{AB\xi \sigma}_{E_i}(\bm{q})$ corresponds to the $p$-wave components, referred to as loop-current fields\cite{Onari_AX2020}.
We note that, if we consider the $D_6$ point group case\cite{HCPo_PRB2019_Faithful} which is another symmetry of MA-TBG, nematic fields appear irrespective of loop-current fields.

\begin{table}[htb]
  \begin{center}
    \begin{tabular}{|c|c|c|c|c|c|} \hline
             & E& C$_{3z}$&  C$_{2y}$ & linear & quadratic \\ \hline 
      $A_1$& 1& 1& -1& & $x^2+y^2$ \\ \hline
      $A_2$& 1& 1& -1& &  \\ \hline
      $E$      & 2& -1& 0& $(x,y)$& $(x^2-y^2, xy)$ \\ \hline
    \end{tabular}
   \caption{The character table of the $D_3$ point group. }
   \label{table:1}
  \end{center}
\end{table}

\section{Effective Action\label{appendix:action}}
Here, we derive the action in Eq. (\ref{eq:action}).
The effective model containing the quadrupole-quadrupole interaction in Eq. (\ref{eq:forward_scattering}) or in Appendix \ref{appendix:interaction} is given by
\begin{eqnarray}
H&=&\frac{1}{N}\sum_{\bm{k}\xi\sigma}
\left(
\begin{array}{cc}
c_{\bm{k}\xi\sigma}^{AB\dagger} & c_{\bm{k}\xi\sigma}^{BA\dagger}  
\end{array}
\right)
\hat{\mathcal{H}}^{ \xi}_{\bm{k}\sigma}
\left(
\begin{array}{c}
c_{\bm{k}\xi\sigma}^{AB} \\
c_{\bm{k}\xi\sigma}^{BA}  
\end{array}
\right) +H_{\rm{int}}
,\nonumber \\ \\
H_{\rm{int}}&=&-g\sum_{\xi \sigma}\sum_{i=1,2}\sum_{\bm{q}}
\Bigl(n^{AB \xi\sigma}_{E_i}(\bm{q})n^{BA\xi\sigma}_{E_i}(-\bm{q})\Bigr),\nonumber \\
\end{eqnarray}
where $g=\frac{2V_{\rm{NN}}}{3}$ is a coupling constant and $\hat{\mathcal{H}}^{\xi}_{\bm{k}\sigma}$ is a $2\times2$ Hamiltonian for each valley $\xi$ and $\sigma$. 
We perform the Hubbard-Stratonovich transformation 
by using the two-component complex field $(\bm{\Psi},\bar{\bm{\Psi}})$, with $\bm{\Psi}=(\Psi_1,\Psi_2)$, $\Psi_i \in \bm{C}$, and $\bar{\bm{\Psi}}=\bm{\Psi}^*$, as follows  
\begin{eqnarray}
S_{\rm int}&=&-\sum_{i\xi \sigma}\int_{q}
\frac{1}{\sqrt{\beta N}}\Bigl(\bar{\Psi}_i(-q)n^{AB\xi\sigma}_{E_i}(q)\nonumber \\ &&+ \Psi_i(-q)n^{BA\xi\sigma}_{E_i}(q)\Bigr)+\frac{1}{g} \sum_{i\xi \sigma}\int_q\Psi_i(-q)\bar{\Psi}_i(q) ,\nonumber \\
&=&\sum_{i\xi \sigma}\int_{q}\int_{k}
\sum_{\alpha \beta}\bar{c}_{k+\frac{q}{2}\xi\sigma}^{\alpha}
\Bigl[V^{i}_{\alpha \beta}(k,q)  \Bigr]
c_{k-\frac{q}{2}\xi\sigma}^{\beta}
\nonumber \\ &&+ \frac{1}{g} \sum_{i\xi \sigma}\int_q \Psi_i(-q)\bar{\Psi}_i(q),
\end{eqnarray}
where $V^i_{\alpha \beta}(k,q)$ is an $(\alpha, \beta)$ component of the matrix $\hat{V}^i(k,q)$. In terms of form factors $E_{\bm{k}}^i$ in Eq. (\ref{eq:form}), it is expressed,
\begin{eqnarray}
\hat{V}^i(k,q)&=&
-\frac{1}{\sqrt{\beta N}}
\left(
\begin{array}{cc}
0 & E^{i*}_{\bm{k}}\bar{\Psi}_i(-q)\\
 E^{i}_{\bm{k}}\Psi_i(-q)& 0
\end{array}
\right),
\end{eqnarray}
with $q=(\bm{q},i\omega_n)$, $k=(\bm{k},i\omega_m)$, where $\sigma_x, \sigma_y$ are Pauli matrices and $\bar{\bm{c}},\bm{c}$ are Grassmannian variables corresponding to creation and annihilation operators.

Next, we divide $\Psi$ into nematic fields $\Phi_i(q)$ and loop-current fields $\Phi^{\prime}_i(q)$, 
where $\Phi_i(q)={\rm Re}\Psi_i(q) \in \mathbb{R}$ and $\Phi_i^{\prime}(q)={\rm Im}\Psi_i(q) \in \mathbb{R}$. In the following calculation, for simplicity, we only consider an electron-nematic order and in this case the matrix $\hat{V}^i(k,q)$ is written as 
\begin{eqnarray}
\hat{V}^i(k,q)&=&
-\frac{\Phi_i(-q)}{\sqrt{\beta N}}
\left(
\begin{array}{cc}
0 & E^{i*}_{\bm{k}}\\
 E^{i}_{\bm{k}}& 0
\end{array}
\right).
\end{eqnarray}
The total action in this system is given by the two-component real field $\bm{\Phi}=(\Phi_1,\Phi_2)$,
\begin{eqnarray}
S_{\rm tot}[\bar{\bm{c}},\bm{c},\bm{\Phi}]&=&S_{\rm 0}[\bar{\bm{c}},\bm{c}]+S_{\rm int}[\bar{\bm{c}},\bm{c},\bm{\Phi}],
\end{eqnarray}
with
\begin{eqnarray}
S_{\rm 0}[\bar{\bm{c}},\bm{c}]
&=&\sum_{i\xi \sigma}\int_{q}\int_{k}
\sum_{\alpha \beta}\bar{c}_{k+\frac{q}{2}\xi\sigma}^{\alpha}
\Bigl[ \bigl(-i\omega_n \delta_{\alpha \beta}+\mathcal{H}^{\xi}_{\bm{k}\sigma, \alpha \beta }\bigr) \nonumber \\&&\times \delta_{q,0}\Bigr]
c_{k-\frac{q}{2}\xi\sigma}^{\beta}.
\end{eqnarray}
After integrating out the electron degrees of freedom, we have an effective action for the nematic field,  
\begin{eqnarray}
S_{\rm eff}[\bm{\Phi}]
&=&\frac{1}{g} \sum_{i\xi \sigma}\int_q\Phi_i(-q)\Phi_i(q) -{\rm Tr}{\rm ln}\Bigl[  \hat{M}_{k+\frac{q}{2}, k-\frac{q}{2}}^{\xi \sigma}\Bigr] , \nonumber \\ \\
\hat{M}_{k+\frac{q}{2}, k-\frac{q}{2}}^{\xi \sigma}&=&
-\hat{G}_0^{-1}\delta_{k+q/2,k-q/2}-\sum_{i=1,2}\hat{V}^i(k,q),\nonumber \\
\label{eq:M}
\end{eqnarray}
where we have introduced the non-interacting Green's function $\hat{G}_0^{-1}(k)=i\omega_n \1-\hat{\mathcal{H}}^{\xi}_{\bm{k}\sigma}$.
This leads to Eq. (\ref{eq:action}).

\section{Ginzburg-Landau Expansion\label{appendix:GLexpansion}}
Here, we derive the GL expansion in Eq. (\ref{eq:GLW_action_sixth}).
For simplicity, we approximate the $2\times 2$ Dirac Hamiltonian with chiral symmetry on the honeycomb lattice,
\begin{eqnarray}
\hat{H}_{\bm{k}}&=&
\left(
\begin{array}{cc}
\mu& \epsilon_{\bm{k}}^*\\
\epsilon_{\bm{k}}&\mu
\end{array}
\right),\\
\hat{U}_{\bm{k}}&=&
\frac{1}{\sqrt{2}}\left(
\begin{array}{cc}
1& 1\\
e^{i\theta_{\bm{k}}}&-e^{i\theta_{\bm{k}}}
\end{array}
\right),
\end{eqnarray}
where a phase factor is introduced as $\theta_{\bm{k}}=\frac{\epsilon_{\bm{k}}}{|\epsilon_{\bm{k}}|}$ with $\epsilon_{\bm{k}}=t(1+e^{-i\bm{k}\cdot\bm{a}_1}+e^{-i\bm{k}\cdot\bm{a}_2})$, the hopping parameter $t$, and the chemical potential $\mu$.  
The band representation of the non-interacting Green's function and the interaction vertex in Eq. (\ref{eq:M})
is given by
\begin{eqnarray}
\hat{G}_0(k)&=&\hat{U}_{\bm{k}}
\left(
\begin{array}{cc}
g^+_{k} & 0\\
0& g^-_{k} 
\end{array}
\right)\hat{U}_{\bm{k}}^{\dagger}, 
\label{eq:free_green}
\\
\hat{V}^i(k,q)&=&
-\frac{1}{\sqrt{\beta N}}
\hat{U}_{\bm{k}}\Bigl\{
\left(
\begin{array}{cc}
1& 0\\
0&-1
\end{array}
\right)
\nonumber \\ &&\times \Bigl[
d^{i}_{\bm{k}}\Phi_i(-q)-p^{i}_{\bm{k}}\Phi^{\prime}_i(-q)
\Bigr] \nonumber \\
&+&\left(
\begin{array}{cc}
0& -i\\
i&0
\end{array}
\right)
\Bigl[
-p^{i}_{\bm{k}}\Phi_i(-q)+d^{i}_{\bm{k}}\Phi^{\prime}_i(-q)
\Bigr]
\Bigr\}\hat{U}_{\bm{k}}^{\dagger}, \nonumber \\
\end{eqnarray}
where $[g^{\pm}_k]^{-1}=i\omega_n\mp|\epsilon_{\bm{k}}|-\mu$ is the electron Green's function, $\omega_n$ is the fermion Matsubara frequency, and we have introduced the $d$- and $p$- wave components of the form factor $E_{\bm{k}}^i$ in Eqs.(\ref{eq:form}),
\begin{eqnarray}
d^i_{\bm{k}}&=&{\rm Re} \Bigl[ E_{\bm{k}}^i e^{-i\theta_{\bm{k}}}\Bigr],\\
p^i_{\bm{k}}&=&{\rm Im} \Bigl[ E_{\bm{k}}^i e^{-i\theta_{\bm{k}}}\Bigr].
\label{eq:d_and_p}
\end{eqnarray}
If the system has space inversion symmetry, the $p$-wave component of the nematic field vanishes.
Now we focus on the $d$-wave component, for which the matrix $\hat{V}^i(k,q)$ is obtained in a diagonal form,
\begin{eqnarray}
\hat{V}^i(k,q)&=&
-\frac{d_{\bm{k}}^i \Phi_i(-q)}{\sqrt{\beta N}}
\hat{U}_{\bm{k}}
\left(
\begin{array}{cc}
1 & 0\\
0&-1  
\end{array}
\right)\hat{U}_{\bm{k}}^{\dagger}, \nonumber \\
&=&\hat{v}^i(k,q)\Phi_i(-q),
\label{eq:vertex}
\end{eqnarray}
where we have introduced the shorthand notation of the interaction vertex $\hat{v}^i(k,q)$.

As described in Appendix \ref{appendix:action}, we have used the effective action in Eq. (\ref{eq:M}).
In terms of the order parameter field $\bm{\Phi}=(\Phi_1,\Phi_2)$, the partition function is expressed in a functional integral form,
\begin{eqnarray}
Z&=&Z_0\int \mathcal{D} \bm{\Phi}e^{-S_{\rm{GL}}[\bm{\Phi}]}, \\
S_{\rm{GL}}[\bm{\Phi}]&=&\sum_{n=1,\cdots,6}S^{(n)}_{\rm{GL}}[\bm{\Phi}].
\end{eqnarray}
We expand the above GL action up to the sixth order terms in the nematic order parameter, by using the following relation: 
${\rm Tr}{\rm ln} M={\rm Tr}{\rm ln} \bigl(-\hat{G}_0^{-1}\bigr)-\sum_{n=1}^{\infty} \frac{1}{n}{\rm Tr} \bigl( \hat{G}_0\hat{V}\bigr)^n$ where $\hat{M}$ is shown in Eq.(\ref{eq:M}).
The first order term of $\bm{\Phi}$ is 
\begin{eqnarray}
{\rm Tr} \bigl(\hat{G}_0\hat{V}\bigr)&=&
\frac{1}{\sqrt{\beta N}}\sum_{i}\int_k 
 \Bigl[\hat{G}_{k}\Bigr]_{\alpha \beta}
 \Bigl[\hat{v}^i(k,q)\Bigr]_{\beta \alpha}\Phi_i(-q), \nonumber \\
\end{eqnarray}
where this integration becomes zero because $\hat{G}_k$ has $C_3$ symmetry.
The second order term is 
\begin{eqnarray}
\frac{1}{2}{\rm Tr} \bigl(\hat{G}_0\hat{V}\bigr)^2
&=&
\frac{1}{2}\int_{q}\sum_{ij} 
\chi_q^{ij}\Phi_i(-q) \Phi_j(q) ,\nonumber \\ \\
\chi_q^{ij}&=&\frac{T}{N} \sum_{\bm{k},i\omega_n}{\rm tr} \Bigl[ \hat{G}_{k+q/2}\hat{v}^i(k,q)\hat{G}_{k-q/2}\hat{v}^j(k,-q)\Bigr]. \nonumber \\
\end{eqnarray}
Thus the Gaussian term is
\begin{eqnarray}
S^{(2)}_{\rm{Gauss}}[\bm{\Phi}]&=&\int_q  \sum_{ij}  \Bigl[\chi_d^{-1}(q)\Bigr]_{ij} \Phi_i(-q)\Phi_j(q),\\
\Bigl[\chi_d^{-1}(q)\Bigr]_{ij}&=&\frac{1}{g}\delta_{ij}-\chi_q^{ij}, \nonumber  \\
&\sim&(r+\xi_0^2\bm{q}^2)\delta_{ij}+\hat{D}_q^{ij},\label{eq:GL_Gauss}
\end{eqnarray}
where $g=\frac{2V_{\rm{NN}}}{3}$ is a coupling constant, the dynamical part of nematic fluctuation is defined as $\hat{D}_q=\hat{\chi}_q-\hat{\chi}_{\bm{q},0}$, $r =1/g-\chi_{q=0}^{11}\propto T_{c0}-T$ measures the distance from the mean-field transition temperature $T_{c0}$, and the mean-field correlation length is $\xi_0$.

In a similar way, we evaluate the coefficients up to sixth order.
GL coefficients $u_{n+1}$ comes from uniform contributions of the $(n+1)$-th order term in the non-interacting Dirac dispersion, 
\begin{eqnarray}
u_{n+1}&=&\frac{T}{N} \sum_{\bm{k},i\omega_n}{\rm tr} \Bigl[ \hat{G}_{k}\hat{v}^{i}(k,0) \Bigr]^{n+1},  \\
&=&\frac{1}{N}\sum_{\bm{k}}  (d^i_{\bm{k}})^{n+1} 
\frac{1}{n!}\Bigl[
\frac{\partial^{n}}{\partial^{n}\epsilon} 
f(\epsilon^{+}_{\bm{k}})+(-1)^{n+1}
\frac{\partial^{n}}{\partial^{n}\epsilon} 
f(\epsilon^{-}_{\bm{k}})\Bigr].\nonumber \\
\label{eq:coefficient}
\end{eqnarray}
where we have used the non-interacting formula in Eqs. (\ref{eq:free_green}) and (\ref{eq:vertex}) 
with $T\sum_{i\omega_n}[g^i(k)]^{n+1}=\frac{1}{n!}\frac{\partial^n}{\partial^n \epsilon} f(\epsilon^i)$ at the second line.
If we treat the impurity effect in a Born approximation, the electron Green's function is evaluated as 
$\hat{G}^{-1}(k)=\hat{G}^{-1}_{0}(k)-\hat{\Sigma}_{\rm{imp}}(k)$ and the self-energy is obtained in Eq. (\ref{eq:born_selfenergy}).

As a consequence, we arrive at the following GL action up to sixth order,
\begin{eqnarray}
S_{\rm{GL}}[\bm{\Phi}]&=&
\int_x \Bigr[\frac{1}{2}r \Phi_+\Phi_-
+\frac{1}{6}u_3(\Phi_+^3+\Phi_-^3)
+\frac{1}{4}u_4 \Phi_+^2\Phi_-^2 \nonumber \\
&+&\frac{1}{10}u_5(\Phi_+^4\Phi_-+\Phi_+\Phi_-^4)
+\frac{1}{6}u_6\Phi_+^3\Phi_-^3
\Bigl],
\end{eqnarray}
where $x=(\bm{r},\tau)$,$\Phi_{\pm}=\Phi_{1}\pm i\Phi_{2}$ and $r=1/g-u_2$ as shown in Eq. (\ref{eq:GLW_action_sixth}).

\section{Electron-phonon coupling\label{appendix:electron-phonon}}

Here, we derive the electron-phonon coupling in Eq. (\ref{eq:electron_phonon}) from a change in the bond length\cite{Suzuura_PRB2002,CastroNeto_RMP2009,Vozmediano_PhysRep2010}.
We assume that the electron-phonon coupling arises from the lattice modulation by phonons, which leads to a change in the nearest-neighbor hopping $t$, 
\begin{eqnarray}
H_{\rm el-ph}&=&\sum_{\bm{\delta}}\bm{g}(\bm{\delta})\sum_{\bm{r}_i,\xi\sigma}
\Bigl[\bm{u}_{\alpha}(\bm{r}_i)-\bm{u}_{\beta}(\bm{r}_i+\bm{\delta}) \Bigr]
\sum_{\alpha}c_{\xi\sigma}^{\alpha\dagger}c_{\xi\sigma}^{\bar \alpha},\nonumber \\  
\end{eqnarray}
where $\bm{u}_{\alpha}(\bm{r}_i)$ is the lattice displacement vector at $\bm{r}_i$, $\bm{\delta}$ is the nearest neighbor lattice vector, $\bm{g}(\bm{\delta})=\nabla t (\bm{\delta})=g_{\rm{nn}}\bm{\delta}_{\rm{nn}}$ with the hopping amplitude $t (\bm{\delta})$ between sites $\bm{r}_i$ and $\bm{r}_i+\bm{\delta}$. The Fourier representation of the electron-phonon coupling is 
\begin{eqnarray}
H_{\rm el-ph}&=&
\sum_{\bm{\tau}_i}\bm{\tau}_i\frac{g_{\rm ph}}{\sqrt{N}}\sum_{\bm{p}\bm{q}}
\Bigl[\bm{u}_{A\bm{q}}-\bm{u}_{B\bm{q}}e^{i\bm{q}\cdot \bm{\tau}_i} \Bigr]\nonumber \\
&\times&\Bigl[c_{A \bm{k}+\bm{q}/2}^{\dagger}c_{B \bm{k}-\bm{q}/2}e^{i\bm{k}\cdot \bm{\tau}_i}\nonumber \\ &&
+c_{B \bm{k}+\bm{q}/2}^{\dagger}c_{A \bm{k}-\bm{q}/2}e^{i(\bm{k}+\bm{q}/2)\cdot \bm{\tau}_i}\Bigr]
,\nonumber \\
&=&  
\frac{g_{\rm ph}}{\sqrt{N}}\sum_{\bm{p}\bm{q}}
\Bigl[\bm{u}_{A\bm{q}}\cdot 
\left(
\begin{array}{c}
E_{\bm{k}-\bm{q}/2}^{1} \\ E_{\bm{k}-\bm{q}/2}^{2}  
\end{array}
\right)
-\bm{u}_{B\bm{q}}\cdot
\left(
\begin{array}{c}
E_{\bm{k}+\bm{q}/2}^{1} \\ E_{\bm{k}+\bm{q}/2}^{2}  
\end{array}
\right)
\Bigr]\nonumber \\ &\times&
c_{A \bm{k}+\bm{q}/2}^{\dagger}c_{B \bm{k}-\bm{q}/2}
\nonumber \\
&+&\frac{g_{\rm ph}}{\sqrt{N}}\sum_{\bm{p}\bm{q}}
\Bigl[\bm{u}_{A\bm{q}}\cdot 
\left(
\begin{array}{c}
E_{\bm{k}+\bm{q}/2}^{1*} \\ E_{\bm{k}+\bm{q}/2}^{2*}  
\end{array}
\right)
-\bm{u}_{B\bm{q}}\cdot
\left(
\begin{array}{c}
E_{\bm{k}-\bm{q}/2}^{1*} \\ E_{\bm{k}-\bm{q}/2}^{2*}  
\end{array}
\right)
\Bigr]\nonumber \\ &\times&
c_{B \bm{k}+\bm{q}/2}^{\dagger}c_{A \bm{k}-\bm{q}/2},
\nonumber \\
&=&\frac{g_{\rm ph}}{\sqrt{N}}\sum_{\bm{p}\bm{q}}
\Bigl[\bm{u}^{\rm OP}_{\bm{q}}\cdot 
\bm{E}_{\bm{k}}+\bm{u}^{\rm AC}_{\bm{q}}\cdot 
\Delta \bm{E}_{\bm{k},\bm{q}}\Bigr]\nonumber \\ &\times&
c_{A \bm{k}+\bm{q}/2}^{\dagger}c_{B \bm{k}-\bm{q}/2} \nonumber\\ 
&+&\frac{g_{\rm ph}}{\sqrt{N}}\sum_{\bm{p}\bm{q}}
\Bigl[\bm{u}^{\rm OP}_{\bm{q}}\cdot 
\bm{E}_{\bm{k}}^*
+\bm{u}^{\rm AC}_{\bm{q}}\cdot 
\Delta \bm{E}_{\bm{k},\bm{q}}^*\Bigr]\nonumber \\ &\times&c_{B \bm{k}+\bm{q}/2}^{\dagger}c_{A \bm{k}-\bm{q}/2}, 
\end{eqnarray}
where we have introduced displacement fields of an optical phonon $\bm{u}^{\rm OP}_{\bm{q}}=\frac{1}{\sqrt{2}}(\bm{u}_{A\bm{q}}-\bm{u}_{B\bm{q}})$ and an acoustic phonon $\bm{u}^{\rm AC}_{\bm{q}}=\frac{1}{\sqrt{2}}(\bm{u}_{A\bm{q}}+\bm{u}_{B\bm{q}})$ in the long-wave length limit.
The vectors $\bm{E}_{\bm{k}}$ and $\Delta\bm{E}_{\bm{k},\bm{q}}$ are obtained from the Taylor expansion for small $\bm{q}$ as follows, 
$\bm{E}_{\bm{k}+\bm{q}/2}-\bm{E}_{\bm{k}-\bm{q}/2}=
\bm{E}_{\bm{k}}+\Delta \bm{E}_{\bm{k},\bm{q}} \cdots $ and $\bm{E}_{\bm{k}}=(E_{\bm{k}}^{1}, E_{\bm{k}}^{2} )$,
\begin{eqnarray}
\Delta \bm{E}_{\bm{k},\bm{q}}&=&
\left(
\begin{array}{c}
-\frac{1}{2} \\-\frac{\sqrt{3}}{2}
\end{array}
\right)e^{i\bm{k}\cdot \bm{a}_1}(i\bm{q}\cdot \bm{a}_1)+
\left(
\begin{array}{c}
-\frac{1}{2} \\ \frac{\sqrt{3}}{2}
\end{array}
\right)e^{i\bm{k}\cdot \bm{a}_2}(i\bm{q}\cdot \bm{a}_2). \nonumber \\  
\end{eqnarray}

Finally, the electron-phonon coupling term for acoustic phonons resulting from the bond-length change is
\begin{eqnarray}
&&S_{\rm el-ph}[\bar{\bm{c}},\bm{c},\tilde{u}_{L},\tilde{u}_{T}]\nonumber \\
&=&\sum_{i\xi \sigma}\int_{q}\int_{k}
\sum_{\alpha \beta}\bar{c}_{k+\frac{q}{2}\xi\sigma}^{\alpha}
\Bigl[ \hat{w}^{\mu}(k,q)  \Bigr]_{\alpha \beta}c_{k-\frac{q}{2}\xi\sigma}^{\beta} \tilde{u}_{\mu}(-q), \nonumber \\
\end{eqnarray}
with the displacement field of acoustic phonons $\bm{u}_{\mu=T,L}=\tilde{u}_{\mu}\hat{\bm{e}}_{\mu}$
where $\hat{\bm{e}}_{T}=(-\sin{\theta_{\bm{q}}}, \cos{\theta_{\bm{q}}})$ and $\hat{\bm{e}}_{L}=(\cos{\theta_{\bm{q}}}, \sin{\theta_{\bm{q}}})$ with $\theta_{\bm{q}}=\tan{^{-1}(q_y/q_x)}$ and
\begin{eqnarray}
\hat{w}_{k,q}^{\mu}&=&
-\frac{g_{\rm ph}}{\sqrt{\beta N}}
\left(
\begin{array}{cc}
0 & \Delta \bm{E}^{*}_{\bm{k},\bm{q}}\cdot \hat{\bm{e}}_{\mu}(-q)\\
\Delta \bm{E}_{\bm{k},\bm{q}}\cdot \hat{\bm{e}}_{\mu}(-q)& 0
\end{array}
\right).\nonumber \\
\end{eqnarray}
After integrating out the electron degrees of freedom, we have a self-energy correction to the phonon action in Eq. (\ref{eq:phonon_full}),
\begin{eqnarray}
\delta K_{\mu, \rm{el-ph}}(q)&=&-\frac{g^2_{\rm ph}}{2\rho}\int_q {\rm tr} \Bigl[ \hat{G}_{k+q/2}\hat{w}^{\mu}_{k,q}\hat{G}_{k-q/2}\hat{w}^{\mu}_{k,-q}\Bigr]. \nonumber \\
\end{eqnarray}
This is shown in Eq. (\ref{eq:electron_phonon}).

\section{Hartree-Fock approximation of a bond-order\label{appendix:hartreefock}}

Here, we derive the mean-field theory of the three-state Potts nematic phase transition following Ref. \onlinecite{Valenzuela_NJP2008} and show the Fermi surface and DOS in Sec. \ref{section:mean-field}.
The effective model containing the quadrupole-quadrupole interaction is shown in Eq. (\ref{eq:forward_scattering}) and in Appendix \ref{appendix:interaction}.
After introducing the mean-field decoupling $n_{\bm{k}\sigma}^{\alpha \beta}=
n_{\bm{k}\sigma}^{\alpha \beta}-\langle n_{\bm{k}\sigma}^{\alpha \beta}\rangle
+\langle n_{\bm{k}\sigma}^{\alpha \beta}\rangle$ 
and ignoring the second order correction $(n_{\bm{k}\sigma}^{\alpha \beta}-\langle n_{\bm{k}\sigma}^{\alpha \beta}\rangle)$, we arrive at 
\begin{eqnarray}
H_{\rm{int}}^{\rm{MF}}&=&
\frac{1}{N}\sum_{\bm{k},\bm{k}^{\prime}}
\Bigl[
f_{\bm{k},\bm{k}^{\prime}} \langle n_{\bm{k}^{\prime}}^{BA}\rangle n_{\bm{k}}^{AB} 
+f_{\bm{k},\bm{k}^{\prime}}^{*}\langle n_{\bm{k}^{\prime}}^{AB}\rangle  n_{\bm{k}}^{BA} 
\Bigr],\nonumber \\
&&-\frac{1}{N}\sum_{\bm{k},\bm{k}^{\prime}}f_{\bm{k},\bm{k}^{\prime}}
 \langle n_{\bm{k}^{\prime}}^{BA}\rangle \langle n_{\bm{k}}^{AB}\rangle,\nonumber \\
&=&
\sum_{\bm{k}}
\left(
\begin{array}{cc}
c_{\bm{k}}^{A\dagger} & c_{\bm{k}}^{B\dagger}  
\end{array}
\right)
\left(
\begin{array}{cc}
0& \Delta_{\bm{k}}^{AB} \\
 \Delta_{\bm{k}}^{BA} & 0 
\end{array}
\right)
\left(
\begin{array}{c}
c_{\bm{k}}^{A} \\
c_{\bm{k}}^{B}  
\end{array}
\right) \nonumber \\&&-\sum_{k}\Delta_{\bm{k}}^{AB}\langle n_{\bm{k}}^{AB} \rangle, 
\end{eqnarray}
with $f_{\bm{k},\bm{k}^{\prime}}=
g(E^{1*}_{\bm{k}}E^1_{\bm{k}^{\prime}}+E^{2*}_{\bm{k}}E^2_{\bm{k}^{\prime}})$, a coupling constant $g=\frac{2}{3}V_{\rm NN}$, the mean-field $\Delta_{\bm{k}}^{AB}=
\frac{1}{N}\sum_{\bm{k}^{\prime}}f_{\bm{k},\bm{k}^{\prime}}\langle n_{\bm{k}^{\prime}}^{BA} \rangle$ and form factors $E_{\bm{k}}^i$ in Eq. (\ref{eq:form}).
The two-component complex order parameter $(\bm{\Psi},\bar{\bm{\Psi}})$ with $\bm{\Psi}=(\Psi_1,\Psi_2)$ contributes to the above mean-field as,  
\begin{eqnarray}
 \Delta_{\bm{k}}^{AB}&=&-\frac{g}{N}\sum_{\bm{k}^{\prime}}\Bigl[E^{1*}_{\bm{k}}E^1_{\bm{k}^{\prime}}\langle n_{\bm{k}^{\prime}}^{BA} \rangle+E^{2*}_{\bm{k}}E^2_{\bm{k}^{\prime}}\langle n_{\bm{k}^{\prime}}^{BA} \rangle \Bigr],\nonumber \\
&=&\Bigl[ E^{1*}_{\bm{k}}\Psi_1+E^{2*}_{\bm{k}}\Psi_2\Bigr], 
\end{eqnarray}
where the order parameters are defined as $\Psi_{1(2)}=-\frac{g}{N}\sum_{\bm{k}}E^{1(2)}_{\bm{k}}\langle n_{\bm{k}}^{BA} \rangle$.
Moreover, the energy shift resulting from the mean-field theory is 
\begin{eqnarray}
-\sum_{k}\Delta_{\bm{k}}^{AB}\langle n_{\bm{k}}^{AB} \rangle&=&
-\sum_{k}\Bigl[ E^{1*}_{\bm{k}}\Psi_1+E^{2*}_{\bm{k}}\Psi_2\Bigr]\langle n_{\bm{k}}^{AB}\rangle,\nonumber \\
&=&\frac{N}{g}\Bigl[ \Psi_1^*\Psi_1+\Psi_2^*\Psi_2\Bigr].
\end{eqnarray}

For example, for a tight-binding model on the honeycomb lattice
\begin{eqnarray}
\hat{\mathcal{H}}_{\bm{k}}^0&=&
\left(
\begin{array}{cc}
0& t(1+e^{-i\bm{k}\cdot\bm{a}_1}+e^{-i\bm{k}\cdot\bm{a}_2}) \\
t(1+e^{i\bm{k}\cdot\bm{a}_1}+e^{i\bm{k}\cdot\bm{a}_2})& 0 
\end{array}
\right),\nonumber \\
\end{eqnarray}
the mean-field term induces the hopping anisotropy as 
\begin{eqnarray}
\hat{\mathcal{H}}_{\bm{k}}^{\rm{MF}}&=&\left(
\begin{array}{cc}
0& E^{1*}_{\bm{k}}\Psi_1+E^{2*}_{\bm{k}}\Psi_2 \\
E^{1}_{\bm{k}}\Psi_1^*+E^{2}_{\bm{k}}\Psi_2^*& 0 
\end{array}
\right).\nonumber \\
\end{eqnarray}
Finally, we obtain the mean-field Hamiltonian, 
$\hat{\mathcal{H}}_{\bm{k}}^{\rm{MF}}$.

\begin{figure}
\includegraphics[width=8cm]{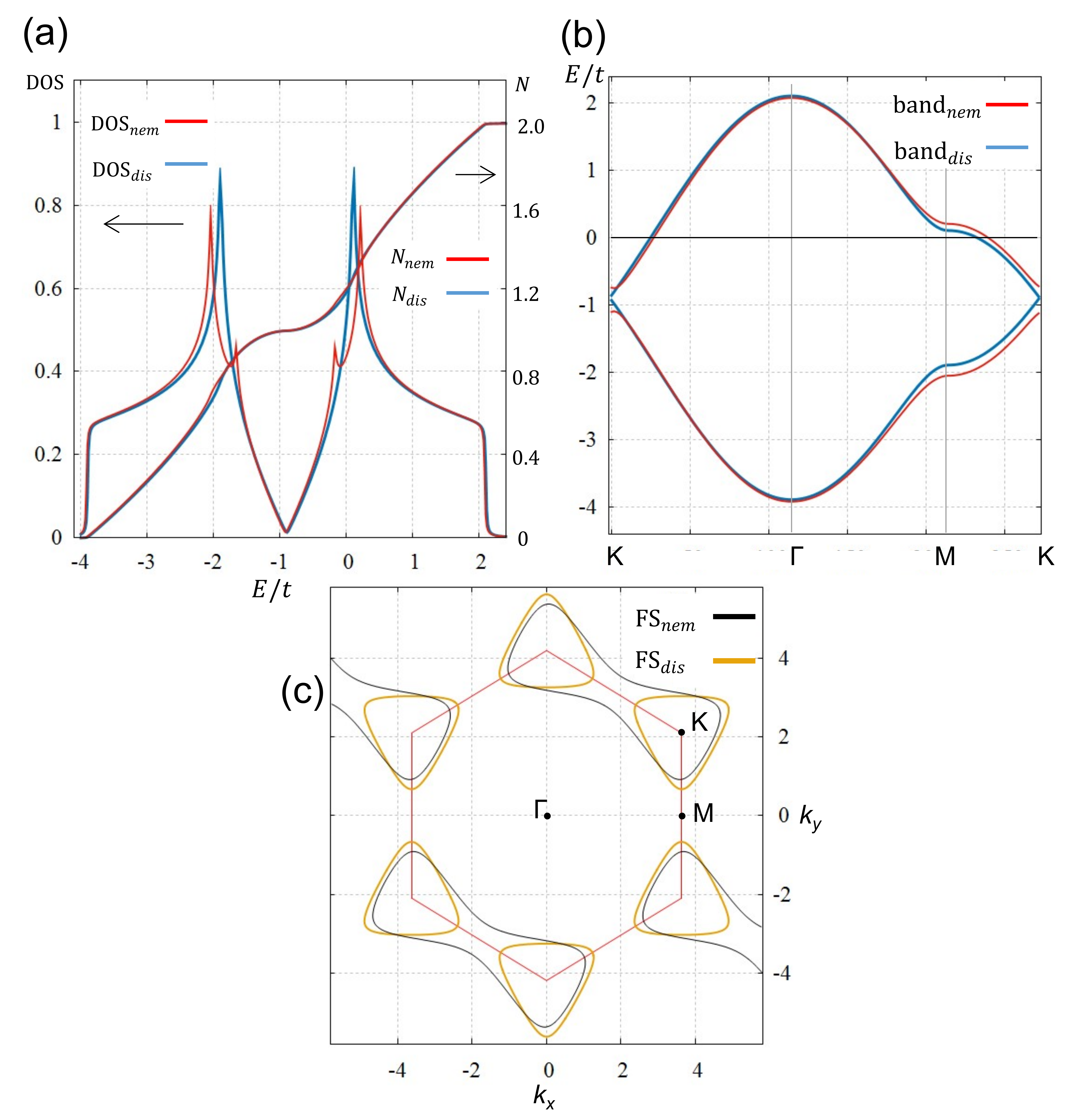}
\caption{
(a)The density of states, and the particle number $N$.
(b)The band structure along the high-symmetric line of the Brillouin zone.
(c)The Fermi surface.
The data are plotted for the disordered phase ($\bm{\Psi}=(0,0)$, $T/t=0.15$) and the nematic phase ($\bm{\Psi}=(0.12,0)$, $T/t=0.05$). 
We use $V_{\rm NN}/t=4.5$, $N=1.2$.
}
\label{fig:band}
\end{figure}

We show numerical results obtained in the Hartree-Fock approximation; The density of states, the band structure and the Fermi surface in the disordered phase and the nematic phase are summarized in Fig.\ref{fig:band}.
In the vicinity of VH filling in Fig.\ref{fig:band}(a), which corresponds to the saddle point of the band in Fig.\ref{fig:band}(b), a finite value of the order parameter yields a deformation of the Fermi surface which breaks the $C_{3z}$ symmetry in Fig.\ref{fig:band}(c).
We note that $N=2$ corresponds to the full filling and $N=1$ corresponds to the charge-neutral point.

\newpage

\bibliography{main}


\end{document}